\documentclass[10pt,twocolumn,letterpaper]{article}

\usepackage{cvpr}      % To produce the REVIEW version
\usepackage[dvipsnames]{xcolor}

\usepackage{pifont}% http://ctan.org/pkg/pifont
\newcommand{\cmark}{\ding{51}}%
\newcommand{\xmark}{\ding{55}}%
\definecolor{cvprblue}{rgb}{0.21,0.49,0.74}
\usepackage[pagebackref,breaklinks,colorlinks,citecolor=cvprblue]{hyperref}

\def\paperID{16685} % *** Enter the Paper ID here
\def\confName{CVPR}
\def\confYear{2024}

\title{CoLay: Controllable Layout Generation through\\Multi-conditional Latent Diffusion}

\author{Chin-Yi Cheng\\
Google Research\\
\and
Ruiqi Gao\\
Google DeepMind\\
\and
Forrest Huang\\
Google Research\\
\and
Yang Li\\
Google Research\\
\and
1600 Amphitheatre Parkway. Mountain View, CA 94043, USA\\
{\tt\small \{cchinyi, ruiqig,  liyang\}@google.com}
}

\begin{document}
% \maketitle
\twocolumn[{%
\renewcommand\twocolumn[1][]{#1}%
\maketitle
\captionsetup{type=figure}
\centerline{\includegraphics[width=.92\textwidth]{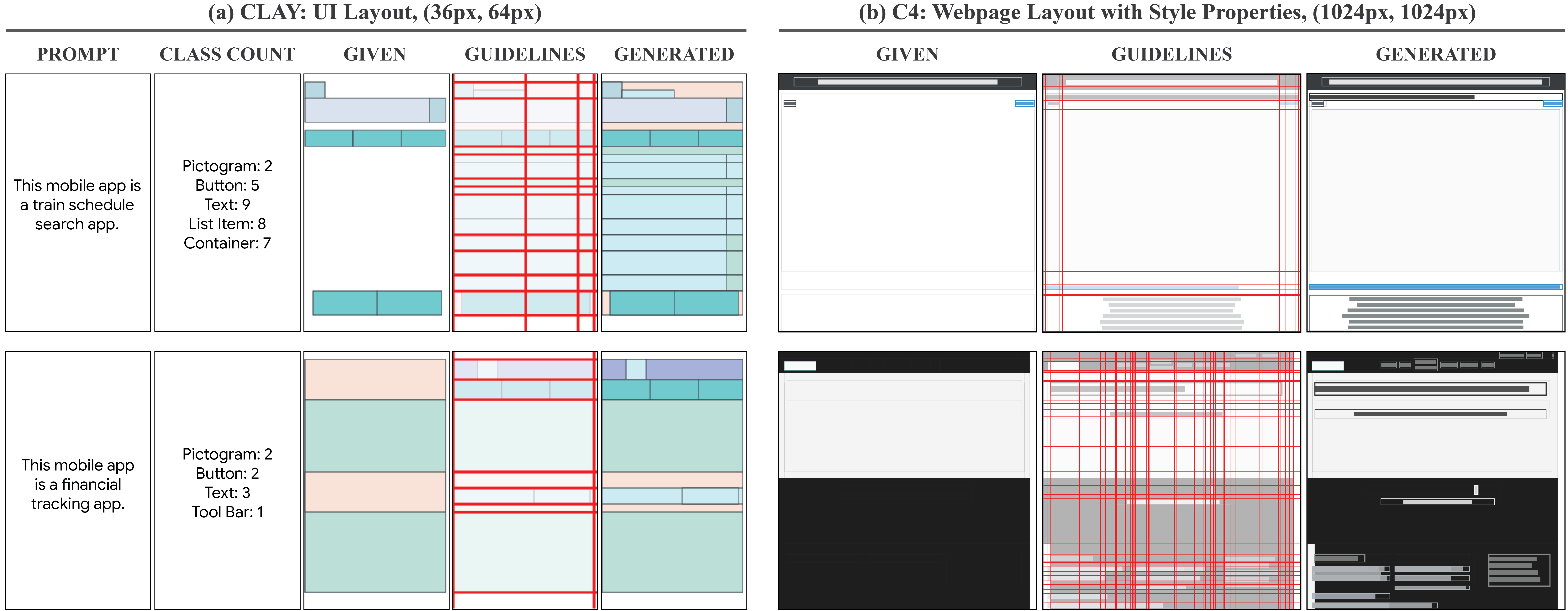}}
\captionof{figure}{Multi-conditional layout generation. Left: UI layouts generated by CoLay trained on the CLAY dataset with four conditions. Right: webpage layouts generated by CoLay trained on the C4 dataset with two conditions.\vspace{2em}}
\label{fig:teaser}
}]
\begin{abstract}
Layout design generation has recently gained significant attention due to its potential applications in various fields, including UI, graphic, and floor plan design. However, existing models face two main challenges that limits their adoption in practice. Firstly, the limited expressiveness of individual condition types used in previous works restricts designers' ability to convey complex design intentions and constraints. Secondly, most existing models focus on generating labels and coordinates, while real layouts contain a range of style properties. To address these limitations, we propose a novel framework, CoLay, that integrates multiple condition types and generates complex layouts with diverse style properties. Our approach outperforms prior works in terms of generation quality and condition satisfaction while empowering users to express their design intents using a flexible combination of modalities, including natural language prompts, layout guidelines, element types, and partially completed designs.
\end{abstract}    
\section{Introduction}
\label{sec:intro}
A layout is a general representation where elements with attributes are placed and arranged within a given boundary. It is widely used in many design and engineering domains, including user interfaces (UI), visual graphics, books, architecture floor plans, and integrated circuits (IC) layouts. A layout design task is a process where designers keep creating and editing elements with the goal to maximize objectives while satisfying constraints. For example, in UI design, designers aim to make the UI layout visually appealing and easy to use, while complying all the functional requirements and design principles. Such a process is usually tedious, iterative, and therefore time-consuming. Therefore, automating the layout design process has become an important research topic.

Many recent works~\cite{layoutganpp, play, lin2023layoutprompter} generate layouts conditionally, where designers can control the generated results by providing a certain type of condition, such as element properties, guideline grids, or text prompts. However, there are two main challenges for existing models to be used in design practice: 
% the limited expressiveness of individual condition types and the lack of style properties in generated layouts.

\begin{enumerate}
    \item \textbf{Limited expressiveness of individual conditions}
    Conditions are usually high-level abstractions of the original designs that more closely align to users' intents, enabling them to guide model generation with little effort The cost of theses abstractions is the limited expressiveness. For example, on the top left of Figure~\ref{fig:teaser}, guidelines are convenient for expressing the desired grids for aligning the elements, but they will never be able to describe the class for the elements. Moreover, a condition can become inefficient and hard to use in certain situations. When using text prompt to express complex space division, for example, we may need to write many sentences instead of simply defining a guideline grid. Therefore, none of the conditions can perfectly handle every possible user intent for layout generation.
    \item \textbf{Lack of style attributes in generated layouts} 
    Most models in existing works generate only box coordinates and class labels. However, a realistic, completed layout should include style properties, such as foreground and background colors, font size, font weight, and text alignment. Although this limitation comes from the datasets, we argue that many existing models may also suffer from the linearly increased token length when generating layouts with multiple style properties.
\end{enumerate}

Our solution to overcome the limited expressiveness on single condition is inspired by real UI design workflows. In the design process, designers always use many representations and tools, such as language, table of specs, design templates, or guideline grids to explore and communicate ideas. They can intuitively select the right tool or combination of tools to address their intent in different situations since they are already familiar with these tools. Therefore, if we provide designers a model with a set of conditions similar to the tools they use daily, they may be able to easily compose these conditions to address complex ideas as needed. Based on this assumption, we propose CoLay, controllable layout generation through multi-conditional latent diffusion, which can be flexibly conditioned on four commonly used conditions: text prompt, class and count, given design, and guidelines, and generate high quality layouts.

Since existing datasets may not contain these conditions, we provide methods to generate and extract conditions from existing datasets, such as prompting LLMs for text summaries of the UIs. We also propose three new metrics to evaluate the relevance between the conditions and the UIs: CycSim for text prompt, C-Usage for class and count, and Design Distance for given design. 

Overcoming the challenge of the lack of style attributes requires a new dataset. Therefore, we leverage C4~\cite{2020t5}, a public dataset of web corpus, and extract the CSS attributes from the rendered view hierarchy of each webpage. With C4, we extend CoLay's capability on generating layouts with more complete visual properties, and show that our latent diffusion approach can scale up well with additional attributes. 

Our experiments show that CoLay outperforms prior works with a significant margin in FID scores, condition metrics, and user study. These improvements are built on top of the existing latent diffusion model, PLay~\cite{play}. We propose a variety of methods to improve it, such as removing the dependency on number of elements, exploring the drop probability for conditions, and sampling with different classifier-free guidance weights for the conditions. Furthermore, we demonstrate a unified experience where users can create and edit layouts step-by-step using the conditions as a toolkit for expressing their thoughts.

In summary, our contributions in CoLay include:
\begin{itemize}
    \item We formulate layout generation as a multi-condition task and include four conditions to express and control complex design intents.
    \item We provide a method to automatically generate prompts for existing layout datasets and introduce a new metric, CycSim, for measuring the text-layout alignment.
    \item We generate realistic layouts with CSS style properties instead of a mock-up with boxes and labels. 
    \item We improve the layout generation quality by a large margin compared to the current SOTA models.
    \item We create a unified, flexible workflow for user to generate and edit layouts by making the model capable of handling any arbitrary subset of the four conditions.
\end{itemize}

%teasor and edit seq, related, user, failure cases, captions.
\section{Related Work}
\label{sec:related}
\subsection{Conditional Layout Generation}
Using generative models for layout generation in vector graphic format has been widely studied ~\cite{layoutgan, layoutvae, layouttransformer}. It is common to represent layouts as a sequence of elements, and therefore, many prior works use sequential models such as transformers as backbone and explore different methods for generation, including autoregressive~\cite{mcl, layouttransformer, blt} and non-autoregressive~\cite{vtn, play}. The best generation quality among recent works are achieved by discrete~\cite{inoue2023layoutdm} and latent diffusion models~\cite{play}. Besides generation quality, it is also important to examine whether the model is capable of generating complex layouts with large number of elements. Most of the existing works use a small number ($<30$) as the maximum number of elements, whereas CoLay can handle complex datasets like C4~\cite{2020t5}.

Conditional generation has also been explored to enable user control, including element attributes and counts~\cite{blt, attribute, layoutganpp}, relationships~\cite{neuraldesign}, images~\cite{canvasvae, cglgan, icvt}, guidelines~\cite{play}, and text prompts~\cite{lin2023layoutprompter}. However, multi-conditional generation with conditions across multiple domains are under-explored. Moreover, although FlexDM~\cite{flexdm} recently explored predicting text style in a layout, it cannot perform well on generating layouts with a large number of elements. Therefore, these limitations make existing models difficult to be used in practice.

\subsection{Conditional Diffusion Models}
Conditional diffusion models~\citep{ho2022cascaded} have demonstrated high effectiveness for controllable generation in various settings, such as image-to-image translation~\citep{saharia2022palette}, text-to-image generation~\citep{ramesh2022hierarchical,saharia2022photorealistic} and text-to-video generation~\citep{ho2022imagen}. By treating any conditional information as an extra input to the model, conditional diffusion models can be trained with the same training objective as their unconditional counterparts. In addition, classifier-free guidance~\citep{ho2022classifier} is an effective technique to further improve the sample quality of conditional diffusion models, where a conditional and an unconditional model are jointly estimated, and the model prediction is modified to be a linear combination of two model outputs.

Conditional diffusion models have also been generalized to multi-conditional settings. For example, ~\cite{nair2022unite, du2023reduce, geffner2022score} study the problem of composing models trained with different single conditions to enable multi-conditional generation. However, these methods either require additional MCMC sampling steps leading to slow generation, or strong assumption that the multiple conditions are conditionally independent at all noise levels. For the conditions we study in this paper, such assumption does not hold. Therefore, we directly feed the model with a random subset of the multiple conditions during training, such that during inference, the model can generate samples given an arbitrary subset of conditions.
\section{Multi-conditinal Layout Generation}
\label{sec:multicond}
\subsection{Datasets}
We conduct experiments for CoLay using two publicly available datasets, CLAY~\cite{clay} and C4~\cite{2020t5}, for UI and webpage layouts respectively. Although CLAY and C4 differ in many aspects such as size, average number of elements, and screen resolution, the main reason we add C4 to our experiments is to test whether the latent diffusion method can be generalized to large number of attributes. Prior works on layout generation generate only class and box coordinates because existing datasets including CLAY, RICO-Semantic~\cite{ricosemantic}, and PublayNet~\cite{publaynet}, only have these two types of attributes. With C4, we can render the webpages as html files and extract the CSS style properties for each element.

In addition, a characteristic common to both CLAY and C4 is that their layouts have hierarchy and order. For example, a container element can contain buttons, images, and other containers. In a sequence of layout elements, if we put the images before their parent container, they will be occluded since the parent container is a larger box that covers all the children elements. Other layout datasets such as RICO-Semantics and PublayNet are much simpler and mostly flat, and therefore, are not included in the experiments.

\begin{table}[t]
\caption{Dataset specifications. Note that the original C4 dataset contains 80 million layouts, and in this work we use a 5 million subset of it. Also, for CLAY we use the same resolution as \cite{play} for fair comparison.}
\label{table:datasets}
\vskip 0.15in
\begin{center}
\begin{small}
\begin{tabular}{l|crrrr}
\toprule
Name & Type & Size & \# Attr. & Avg. \# N & Res. in px\\
\midrule
CLAY & UI & 50K & 5 & 19.62 & 36x64\\
C4 & Web & 5M & 16 & 37.71 & 1024x1024\\
\bottomrule
\end{tabular}
\end{small}
\end{center}
\vskip -0.1in
\end{table}

\subsection{Layout and Conditions}
We formulate the layout as a sequence of elements: $E=\{e^1, e^2, ..., e^N\}$, and each element is composed by its attributes, $e^n = \{{a_1}^n, {a_2}^n, ..., {a_D}^n\}$, where the number of attributes $D$ varies by dataset. In CoLay, we fix $N=64$, by padding $E$ with invalid elements $e_\text{invalid}$ and discarding the layouts with $N>64$. Following prior works~\cite{play, skexgen, layouttransformer}, each attribute is converted to a one-hot vector, so we can obtain $E \in \{0, 1\}^{N \times d_{total}}$, where $d_{total} = \sum_{i=1}^D d_i$. The dimension of the one-hot vector $d_i$ for each attribute $a_i$ can be found in Table~\ref{table:attributes}.

\begin{table}[t]
\caption{Attributes used in CLAY and C4. The attributes listed in the same row share the dimension. For example, the $(r,g,b,a)$ colors have the same range of values from 0 to 255. Also, the attribute value of the invalid element is equal to $d+1$.}
\label{table:attributes}
\vskip 0.15in
\begin{center}
\begin{small}
\begin{tabular}{l|c|rr}
\toprule
Type & Attributes & d-CLAY & d-C4\\
\midrule
Class & $class$ & 24 & 4\\
Coords. & $x_{min}, y_{min}, x_{max}, y_{max}$ & 64 & 1024\\
Foreground & $(r, g, b, a)$ & N/A & 256\\
Background & $(r, g, b, a)$ & N/A & 256\\
Font Size & $font$-$size$ & N/A & 128\\
Font Weight & $font$-$weight$ & N/A & 9\\
Alignment & $alignment$ & N/A & 9\\

\bottomrule
\end{tabular}
\end{small}
\end{center}
\vskip -0.1in
\end{table}

% Other than the class label and box coordinates, $x_{min}, y_{min}, x_{max}, y_{max}$, the C4 dataset contains style attributes including foreground and background colors in $(r, g, b, a)$, font size, font weight, and text alignment.  
%shaped: order container 

\begin{figure*}[t]
  \centering
   \includegraphics[width=0.85\linewidth]{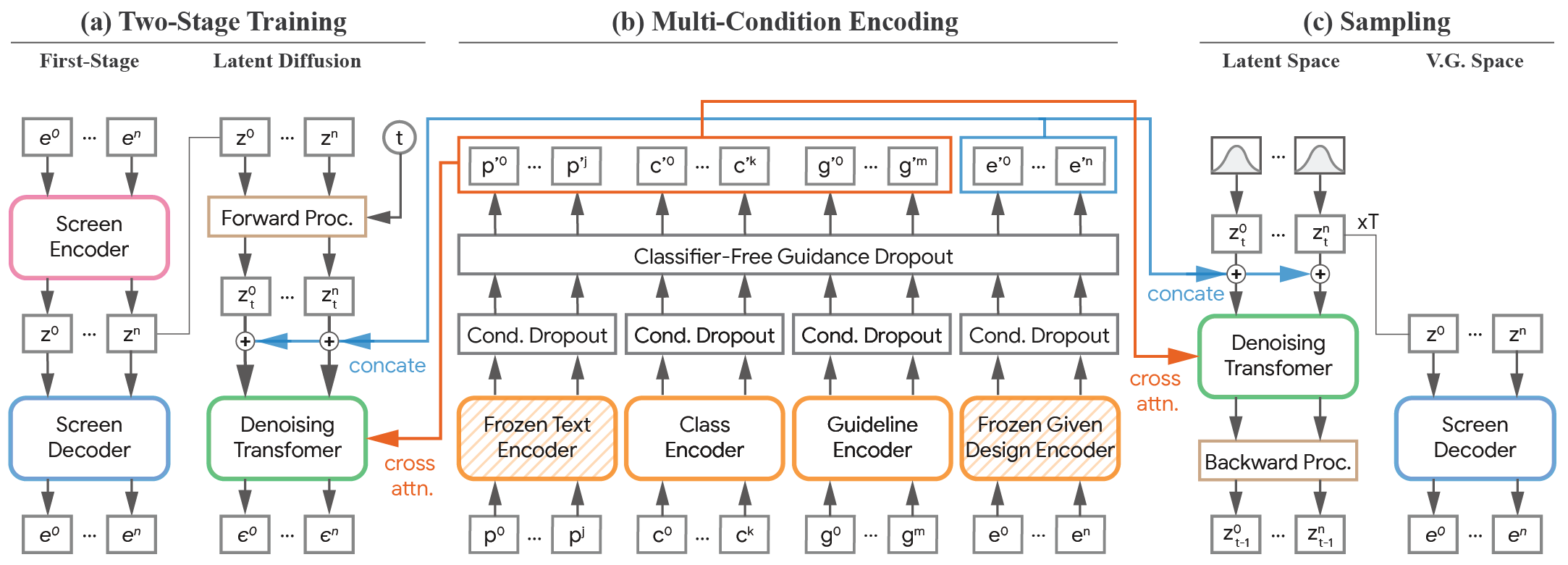}

   \caption{Model architecture. (a) The VAE model is trained to convert the layouts between vector graphic and latent space, and the denoising network is trained on the encoded latent representations. (b) Each condition is encoded by a specific encoder and passed through a dropping mechanism. (c) During sampling, the denoising network will generate $\hat{z_0}$, and the decoder will decode it back to layout.}
   \label{fig:archi}
\end{figure*}

\textbf{Text Prompt}:
None of the CLAY and C4 dataset have the corresponding text prompts for the layouts. To obtain the prompt-layout pairs, we feed view-hierarchy of each sample in these datasets to a PaLM-2-Bison~\cite{palm2}-based LLM, and ask it to generate a paragraph of screen summary $P' = \{p^1, p^2, ..., p^J\}$, where $p^j$ represents each word token in the summary. The view hierarchies are first converted into a simplified HTML format similar to prior work~\cite{bryan}, and sent along with instructions for the LLM to summarize the UI layout as prompts for the LLM. These summaries are then treated as the text prompt and paired with layout $E$ for training. Please refer to the appendix for the LLM task formulation.
% The generated summary from the LLM is then first randomly sampled at the sentence level to obtain $P \subset P'$, then treated as the text prompt and paired with layout $E$ for training. 

\textbf{Class and Count}:
The class and count condition is defined as $C = \{c^1, c^2, ..., c^K\}$, and $K$ is the number of classes. $C$ is obtained by computing the count of each class in a layout. For example, $\{3, 0,..., 0\}$ means that there are three elements in the layout, and all of them belong to the first class.

\textbf{Given Design}:
The given design $E^\prime$ is a partially completed layout. In \cite{play}, this task is formulated as inpainting, where some elements are masked out to be recreated. However, in reality, users may not have the original or desired layout to start with, so they cannot create the corresponding masks. Therefore, instead of using masks, we extract the random subset from the original layout $E$ and pad the missing ones as invalid elements at the end of the sequence.

\textbf{Guidelines}:
We follow PLay and represent the guidelines of a layout as $G = \{g^1, g^2, ..., g^M\}$, and each guideline is defined by concatenating one-hot vector of its axis and position. The guidelines are extracted from the layout by projecting the four coordinates of each box to the $X$ and $Y$ axes and removing the duplication.

\subsection{Metrics}
The quality of conditional layout generation is evaluated by the sample quality and level of satisfaction of the conditions. In line with recent layout generation works \cite{play, inoue2023layoutdm, houseganp}, we choose Fréchet Inception Distance (FID)~\cite{fid} as the main metric for evaluating sample quality. We also define a metric for each of the four conditions used in CoLay to evaluate their levels of satisfaction.

\textbf{FID}: FID measures the distance between the real and generated image distributions by encoding the image samples using a pre-trained Inception \cite{inception} model and computing the Fréchet Distance between the embeddings. For all the experiments, we sample $1024$ samples from both real and generated layouts and render them as images. For CLAY, we use the same color legend as \cite{play}; and for C4, since the layouts already contain style properties, we directly render these properties with dummy text.

\textbf{CycSim for Text Prompt}: Although CLIP score has been commonly used to measure the text-image alignment, it is not applicable to layouts since the CLIP model is not trained on the pairs of layout rendering and prompt. Therefore, we propose CycSim, cycle similarity as the metric to measure the prompt-layout alignment. The idea of CycSim is inspired by the cycle-consistency proposed by Cycle-GAN~\cite{cycle}. The intuition is that the layouts with prompts that have similar meanings may be visually similar, and the prompts with visually similar layouts may also share similar meanings. Based on this intuition, we develop two scores, CycSim-P and CycSim-L.

\begin{itemize}
    \item \textbf{CycSim-P}: For a prompt $P$, we first generate the layout $E_P$ and find $E_P$'s top-k visually similar layouts $\mathbf{E} = \{E_{top1}, E_{top2}, ..., E_{topk}\}$ from the dataset samples. Then we define CycSim-P as the average sentence similarity between $P$ and the corresponding prompts of $\mathbf{E}$.
    \item \textbf{CycSim-L}: For a generated layout $E_P$, we first use the input prompt $P$ to find its top-k similar prompts $\mathbf{P} = \{P_{top1}, P_{top2}, ..., P_{topk}\}$. Then CycSim-L is defined by computing the average visual similarity between $E_P$ and the corresponding layouts of $\mathbf{P}$.
\end{itemize}

We define the visual similarity between a pair of layouts as their cosine similarity of the Inception embeddings. For sentence similarity between two prompts, we encode them using Universal Sentence Encoder~\cite{use} and compute the cosine similarity. We set $k=100$ in the experiments.

%In Cycle-GAN, the loss is computed by the $L2$ norm between image $A$ and $A''$, which is obtained by converting image $A$ to $B'$ and converting $B'$ back to $A''$, and vice versa for image $B$ and $B''$.

%Supp: We tried to fine-tune CLIP but failed.

\textbf{C-Usage for Class and Count}:
To evaluate the class and count usage, we first compute the $L1$ distance $\delta(C, C^\prime)$ between the input class count $C$ and the class count of the generated layout $C^\prime$. We set the distance between $c_k$ and $c_k^\prime$ to zero if $c_k <= c_k^\prime$. The C-Usage is then computed by averaging $1-(\delta(C, C^\prime)/\sum C)$ across generated samples. The reason we clip the distance is that, in practice, the model still has to generate necessary elements even when the user did not specify the count in a class or specified less than needed. For example, if the user specified four images, it does not mean buttons and text are not needed for a legit layout.
%compute C-Usage-Exact

\textbf{Design Distance for Given Design}:
To check whether the given design $E^\prime$ is fully used in the final layout $E$, we define Design Distance as the average of the one-way Chamfer Distance from $E^\prime$ to $E$ across the generated samples. We do not need the other direction of the Chamfer Distance because $E$ does not have to be contained in $E^\prime$.

\textbf{G-Usage for Guidelines}:
We use the same definition as PLay~\cite{play} for G-Usage, which is defined by $|G^{inter}|/|G|$, where $G^{inter}$ is the intersection of the input guidelines $G$ and the guidelines $G^\prime$ extracted from the generated layout. Similar to C-Usage, $G^\prime$ can contain more guidelines than $G$ since the model may need to place elements at positions that are not specified by $G$.
\section{Architecture}
\label{sec:archi}
\subsection{Multi-conditional Latent Diffusion}
\label{mcldm}
% benefit of latent diffusion: continuous, compress
We follow PLay \cite{play} and use a transformer-based latent diffusion model for layout generation. The goal is to generate layouts in the continuous, compact latent space and decode the generated latent vectors back to the vector graphic space. The model is composed of three major components: a variational auto-encoder (VAE), a multi-condition encoder, and a denoising network, as shown in Figure~\ref{fig:archi}. The training process has two stages. In the first stage, we train the VAE to map the layouts between the continuous latent space and the discrete vector graphic space, and then we train the multi-condition encoder and denoising network together for generation. 

A major difference from PLay \cite{play} is that we removed the dependency on the number of elements $N$. In \cite{play}, the denoising network is conditioned on $t$, $G$, and $N$, causing a severe issue: the model is not learning how to generate the proper number of elements to fulfill the conditions. For example, a complex guideline grid implies that more elements are needed in the layout, and vice versa. To resolve this, we include the invalid elements as part of the layout. In this way, the model always generate $N = 64$ elements and learns to arrange valid and invalid elements based on the conditions. We will see the improvement in Sec.~\ref{sec:exp}.

\subsubsection{First-stage VAE}
The encoder $\mathcal{E}(E)$ of the VAE encodes the layout elements $E \in \mathbb{R}^{N \times d_{total}}$ as the latent vector $z \in \mathbb{R}^{N \times \hat{d}}$, and the decoder $\mathcal{D}(z)$ decodes $z$ back to $E$. Both $\mathcal{E}(E)$ and $\mathcal{D}(z)$ are non-autoregressive transformer encoders. Following the prior works \cite{play, vtn}, no positional encoding is used for the VAE since the coordinates already provide explicit positional information. We train the first-stage VAE by augmenting $E$ with its random subsets $E^\prime$, so that the encoder can be versatile enough to handle both the complete layouts and the given, partially completed layouts. 

Different from \cite{play}, which claims that VAE mainly serves the purpose of mapping the discrete space to continuous space for continuous diffusion process, one of the important findings in CoLay is that, the VAE also compresses information by shortening the token length. In many existing works on layout generation, including the ones using discrete diffusion process, such as LayoutDM \cite{inoue2023layoutdm}, each token represents a property instead of an element, and therefore, the length of the sequence is $N \times D$ instead of $N$. In \ref{sec:exp}, we will show existing work struggles when $N$ is large. Moreover, when generating properties beyond label and box coordinates, such as color and font size, the property-based token length will grow linearly with $D$ and make the task more difficult to learn, where the element-based token length will remain the same.

%%%%Supp: Loss is applied to label only for invalid elements 

\subsubsection{Multi-condition Encoder}
\label{multi-cond}
We design the multi-condition encoder, $\hat{P}, \hat{C}, \hat{E}, \hat{G} = \tau_\psi(P, C, E^\prime, G)$, as a set of condition encoders with a hierarchical dropout mechanism. During training, all the conditions is randomly dropped together with probability $p_{cfg}$ for classifier-free guidance, and then each condition is dropped independently with probability $p_{cond}$. There are three advantages of this dropout mechanism. First, as mentioned in \cite{controlnet}, when learning on one of conditions, dropout prevents the model from extracting information from other conditions. Secondly, it augments the dataset by providing more conditions-layout pairs. Lastly, and most importantly, it allows the model to learn on random combinations of conditions and provides flexible options to the user during generation. For example, if the model is trained on four conditions, users can control the layout generation by providing any subset of these four conditions. 

\textbf{Condition Sampling}: We further augment the condition-layout pairs by sampling random subsets of each condition during training. As discussed in \cite{play}, sampling random subsets of the condition not only augment the dataset, but also provides flexibility for the user to control the model. Using guideline condition as an example, users can provide a complex grid of guidelines when they have to follow many design rules, or simply draw a couple of guidelines when they only have a rough idea. The specifications and sampling method for each encoder can be found below.

\begin{itemize}

    \item \textbf{Text Prompt Encoder:} The encoded text prompt $\hat{P} = (\hat{P_{seq}}, \hat{P_{pool}})$ is obtained from a pre-trained BERT model \cite{devlin2018bert}, where $\hat{P_{seq}}$ is the sequential embedding, and $\hat{P_{pool}}$ is the pooled embedding. Since the text prompt $P$ is composed by multiple sentences that describes different parts or features of the layout, we can sample random subsets of the sentences from $P$ to simulate the situations where users provide different granularity and details of text prompts.

    \item \textbf{Class and Count Encoder:} The encoder of the class and count condition is a two-layer transformer encoder. In training we randomly sample the classes by replacing zeros as counts of the unused classes.

    \item \textbf{Given Design Encoder:} We reuse the trained first-stage encoder $\mathcal{E}(E)$ to encode the given design condition, as its representation $E^\prime$ is the same as layout $E$. Similar to the class and count condition, the given design conditions are generated by extracting random subsets from $E$ during training, and therefore, additional sampling is not required.

    \item \textbf{Guideline Encoder:} The guideline condition is encoded using a two-layer transformer encoder. In PLay, the probability $p^m$ to sample a guideline $g^m$ is determined by its weight $h$, which is the total length of box edges overlapping with this guideline, normalized by the sum of weights of all guidelines. Since the same sum of weights can be achieved by many guidelines with small weights or few guidelines with large weights, we normalize the weight by the averaged weight over all guidelines in the layout and reformulate the probability as:
    \begin{equation}
    p^m = (p_{base})^{\Bar{h} / h}
    \end{equation}    
    The sampled guidelines from $X$ and $Y$ axes are concatenated and then encoded by the guideline encoder. 
    %Move to supp. However, the disadvantage of this method is that the same sum of weights can be achieved by many guidelines with small weights or few guidelines with large weights, 
    
\end{itemize}

All the encoded conditions, $\hat{P}, \hat{C}, \hat{E}$, and $\hat{G}$, have the same embedding dimension as the query dimension of the denoising network, so that they can be either injected into the transformer layers or concatenated for joint cross-attension.

%%%%cond Sampling methods?
%%%%

\subsubsection{Denoising Network}

%%%%all 64

We also use a non-autoregressive transformer encoder as the denoising network. It predicts the noise $\epsilon$ given the latent representation $z_t$ and conditioned on the encoded conditions $\hat{P}, \hat{C}, \hat{E}, \hat{G}$, and time step $t$. The loss function is defined using the simplified MSE loss proposed by \cite{ddpm}:
\begin{equation}
\mathcal{L} := \\ \mathbb{E}_{t, \epsilon \sim \mathcal{N}(0, 1)}\left[ \parallel \epsilon - \epsilon_\theta(z_t, \hat{P}, \hat{C}, \hat{E}, \hat{G}, t)\parallel^2 \right]   
\end{equation}

We inject the conditions into the transformer layers in three ways: direct concatenation, feature-wise affine, and cross attention. For conditions that have element-wise correspondence, such as the given design $E^\prime$, we concatenate them with $z_t$ along the element axis; for conditions that have global effects, such as $t$ and the pooled prompt embedding $P_{pool}$, we use a feature-wise affine layer \cite{film} to modulate the features; and lastly, for other types of conditions such as $P_{seq}$, we follow \cite{imagen} to concatenate the conditions with $z_t$ as key and value for cross-attention. We also discovered that the model with a joint cross-attention layer that concatenates all the encoded conditions with $z_t$ outperforms the model with separate cross-attention layer for each condition.

%N is not a condition that user can intuitively set or control.

\subsection{Sampling with Multiple Guidance Weights}

% In sampling, we first initialize $z_T$ and iteratively denoise it to $z_0$ using the denoising network with classifier-free guidance. When only a single condition $G$ is given, the process can be defined as follow:
% \begin{equation}
% \begin{split}
%     \hat{\epsilon}_\theta(z_t, \tau_\psi(\phi, \phi, \phi, G), t) &= (1+w_G)\epsilon_\theta(z_t,  \tau_\psi(\phi, \phi, \phi, G), t) \\ &- w_G \epsilon_\theta(z_t,  \tau_\psi(\phi, \phi, \phi, \phi), t) 
% \end{split}
% \end{equation}

In sampling, we first initialize $z_T$, and iteratively denoise it to $\hat{z_0}$ using the denoising network. We apply classifier-free guidance~\citep{ho2022classifier} to enhance the sample quality. A naive approach is using the same guidance weight for multiple conditions. However, given the different modalities of the conditions in our case, the ideal guidance weight for each condition may differ. Therefore, we propose an approach to handle multiple guidance weights, with the minimum number of functional evaluation calls. We describe it in a more general case: assume the model is conditioned on $L$ conditions $\{\mathcal{C}_1, \mathcal{C}_2, ..., \mathcal{C}_L\}$, and the corresponding guidance weights are monotonically increasing: $w_1 \leq w_2 \leq ... \leq w_L$. Then the modified noise prediction is given by:
\begin{equation}
\begin{split}
    & \hat{\epsilon}_\theta(z_t, \mathcal{C}_1,...,\mathcal{C}_L, t) 
    \\ = & (1 + w_1) \epsilon_\theta(z_t, \mathcal{C}_1, ..., \mathcal{C}_L, t) - w_1 \epsilon_\theta(z_t, t) \\ 
    + & \sum_{i=2}^L(w_i - w_{i-1}) \left(\epsilon_\theta(z_t, \mathcal{C}_i, ..., \mathcal{C}_L, t) - \epsilon_\theta(z_t, t)\right).
\end{split}
\end{equation}

Finally, $\hat{z_0}$ is decoded back to the vector graphic domain using the first-stage decoder $\hat{E_0} = \mathcal{D}(\hat{z_0})$. 
\section{Experiments}
\label{sec:exp}

\begin{table*}[t]
\caption{Ablation and baseline comparisons. All models are trained on CLAY with guideline condition only. GL-Sample: applying new guideline sampling method; B-1024: using a large batch size; Gen-All: generating invalid tokens together; Sort: sorting generated layout.}
\label{table:ablation}
\vskip 0.15in
\begin{center}
\begin{small}
% \small
% \begin{sc}
\begin{tabular}{l|c|cccc|cc}
\toprule
Model & N & GL-Sample & B-1024 & Gen-All & Sort & FID $\downarrow$ & G-Usage $\uparrow$\\
\midrule
LayoutDM & 64 & \cmark & & N/A & N/A & 89.25 & \textbf{0.997}\\
PLay & 64 & & & & & 11.41 & 0.962\\ 
& 64 & \cmark & & & & 11.28 & 0.978\\
& 64 & \cmark & \cmark & & & 10.60 & 0.972\\
& 64 & \cmark & \cmark & \cmark & & 8.86 & 0.979\\
CoLay & 64 & \cmark & \cmark & \cmark & \cmark & \textbf{8.77} & 0.979\\
\midrule
LayoutDM & 25 & \cmark & & N/A & N/A & 11.88 & 0.986\\
CoLay & 25 & \cmark & \cmark & \cmark & \cmark & \textbf{6.71} & \textbf{0.988}\\

\bottomrule
\end{tabular}
% \end{sc}
\end{small}
\end{center}
\vskip -0.1in
\end{table*}

\begin{table*}[t]
\caption{Quantitative results on multi-conditional models trained on CLAY. Each row represents a model trained with a specific combination of conditions. The models are evaluated by FID scores and metrics that measure the level of satisfaction on conditions.}
\label{table:mc}
\vskip 0.15in
\begin{center}
\begin{small}
% \small
% \begin{sc}
\begin{tabular}{cccc|cccccc}
\toprule
\multicolumn{4}{c|}{Conditions} & \multicolumn{6}{|c}{Metrics} \\
\cmidrule(r){1-4}\cmidrule(r){5-10}
$G$ & $E^\prime$ & $C$ & $P$ & FID $\downarrow$ & G-Usage $\uparrow$ & Given-Dist. $\downarrow$ & C-Usage $\uparrow$ & CycSim-L $\uparrow$ & CycSim-P $\uparrow$\\
\midrule
\cmark & \cmark & \cmark & \cmark & 6.23 & 0.984 & 0.031 & 0.960 & 0.774 & 0.503\\ 
\cmark & \cmark & \cmark & & 6.94 & 0.986 & 0.030 & 0.961 & \xmark & \xmark\\
\cmark & \cmark & & & 7.23 & 0.986 & 0.031 & \xmark & \xmark & \xmark\\
\cmark & & & & 8.77 & 0.979 & \xmark & \xmark & \xmark & \xmark\\
\cmark & & \cmark & \cmark & 8.22 & 0.982 & \xmark & 0.940 & 0.776 & 0.504\\
\cmark & & & \cmark & 8.42 & 0.979 & \xmark & \xmark & 0.775 & 0.503\\
& & \cmark & \cmark & 8.54 & \xmark  & \xmark & 0.952 & 0.773 & 0.504\\
 & & & \cmark & 8.93 & \xmark & \xmark & \xmark & 0.779 & 0.506\\
\bottomrule
\end{tabular}
% \end{sc}
\end{small}
\end{center}
\vskip -0.1in
\end{table*}

\begin{table*}[t]
\caption{Quantitative Results on the C4 dataset. Similar to Table~\ref{table:mc}, each row illustrates the performance of a model.}
\label{table:c4}
\vskip 0.15in
\begin{center}
\begin{small}
% \small
% \begin{sc}
\begin{tabular}{l|r|cccccc}
\toprule
Model & Cond. & FID & G-Usage & Given-Dist. & C-Usage & CycSim-P & CycSim-L\\
\midrule
PLay & $(C, E^\prime, G)$ & 26.27 & 0.823 & 0.003 & 0.952 & \xmark & \xmark\\
CoLay & $(C, E^\prime, G)$ & 12.45 & 0.895 & 0.002 & 0.974 & \xmark & \xmark\\
CoLay & $(P, C, E^\prime, G)$ & \textbf{11.21} & \textbf{0.905} & \textbf{0.002} & \textbf{0.978} & \textbf{0.534} & \textbf{0.750}\\
\bottomrule
\end{tabular}
% \end{sc}
\end{small}
\end{center}
\vskip -0.1in
\end{table*}

\subsection{Comparison and Ablations}
In this section, we compare CoLay with recent works and report the ablation study under different settings, datasets, and condition combinations. We select PLay~\cite{play} and LayoutDM~\cite{inoue2023layoutdm} as baseline models for comparison because they represent the best latent and discrete diffusion models for layout generation, respectively. Since LayoutDM is not a conditional model, we re-implement it with the guideline condition and classifier-free guidance, while keeping other parameters the same. We also use the same transformer encoder as the denoising network for CoLay, PLay, and LayoutDM. Please refer to Appendix for implementation details.

From Table~\ref{table:ablation} we can find that LayoutDM cannot perform reasonably when $N=64$, and it under-performs CoLay when $N=25$. These results verify one of the main advantages of using latent diffusion for layout generation: compressing the sequence length from $N \times D$ to $N$. The encoder of our first-stage VAE compresses all the attributes in each element as a single vector, and therefore, the token length can remain the same, making the denoising task less difficult compared to discrete diffusion models. We can also observe that CoLay outperforms PLay by a large margin after applying several changes: the new guideline sampling method introduced in~\ref{multi-cond} improves the G-Usage; using larger batch size and sorting the generated layout by their lexicographical order both improve the FID; and lastly, as discussed in~\ref{mcldm}, generating with invalid elements boosts the FID score significantly. 

In Table~\ref{table:mc} and Table~\ref{table:c4}, we investigate how CoLay performs on CLAY and C4, respectively, when training with different combinations of conditions. It is reasonable that the FID scores drop when adding more conditions since more information is provided to the model. We can also observe consistent numbers on the metrics for condition satisfactory, showing no negative impact among the added conditions. It is worth mentioning that CoLay reduces the FID score on C4 by half compared to PLay, demonstrating its robustness on a dataset that has an order of magnitude higher complexity. 

As we want the model trained on four conditions to be capable of handling every combination of conditions during inference, we investigate how the condition drop probability $p_{cond}$ and guidance weights affect generation quality. Table~\ref{table:cfg} shows that the optimal $p_{cond}=0.5$ as it uniformly samples different combinations of conditions during training. Also, with a grid search on the guidance weights for each combination of conditions, we improve the FID significantly compared to a uniform weight. Nevertheless, there are some noticeable gaps in FID scores between the model trained on all conditions and models specialized on a specific combination, we leave the further improvement of all-conditional model for future research.

\begin{table}[t]
\caption{Top: FID scores on different values of condition drop probability. Bottom: FID scores on different combinations of conditions. Single: using a single guidance weight. Multi: using different weights for each condition and search for optimal weights. Specialized: trained on the specific combination.}
\label{table:cfg}
\vskip 0.15in
\begin{center}
\begin{small}
% \small
% \begin{sc}
\begin{tabular}{cccc|c|ccc}
\toprule
\multicolumn{4}{c|}{Combinations} & \multicolumn{1}{c|}{$p_{cond}$} & \multicolumn{3}{|c}{FID} \\
\cmidrule(r){1-4}\cmidrule(r){5-5}\cmidrule(r){6-8}
$G$ & $E^\prime$ & $C$ & $P$ & - & Single & Multi & Specialized\\
\midrule
\cmark & \cmark & \cmark & \cmark & 0.00 & 8.98 & & \\
\cmark & \cmark & \cmark & \cmark & 0.25 & 6.88 & & \\
\cmark & \cmark & \cmark & \cmark & 0.50 & 6.54 & & \\
\cmark & \cmark & \cmark & \cmark & 0.75 & 6.66 & & \\
\midrule
\cmark & \cmark & \cmark & \cmark & 0.50 & 6.54 & \textbf{6.23} & \textbf{6.23}\\
\cmark & \cmark & \cmark & & 0.50 & 6.85 & \textbf{6.62} & 6.94\\
\cmark & \cmark & & & 0.50 & 7.37 & 7.28 & \textbf{7.23}\\
\cmark & & & & 0.50 & 9.76 & 9.69 & \textbf{8.77}\\
\bottomrule
\end{tabular}
% \end{sc}
\end{small}
\end{center}
\vskip -0.1in
\end{table}

\subsection{Multi-conditional Generation and Editing}

\begin{figure*}[t]
  \centering
   \includegraphics[width=0.95\linewidth]{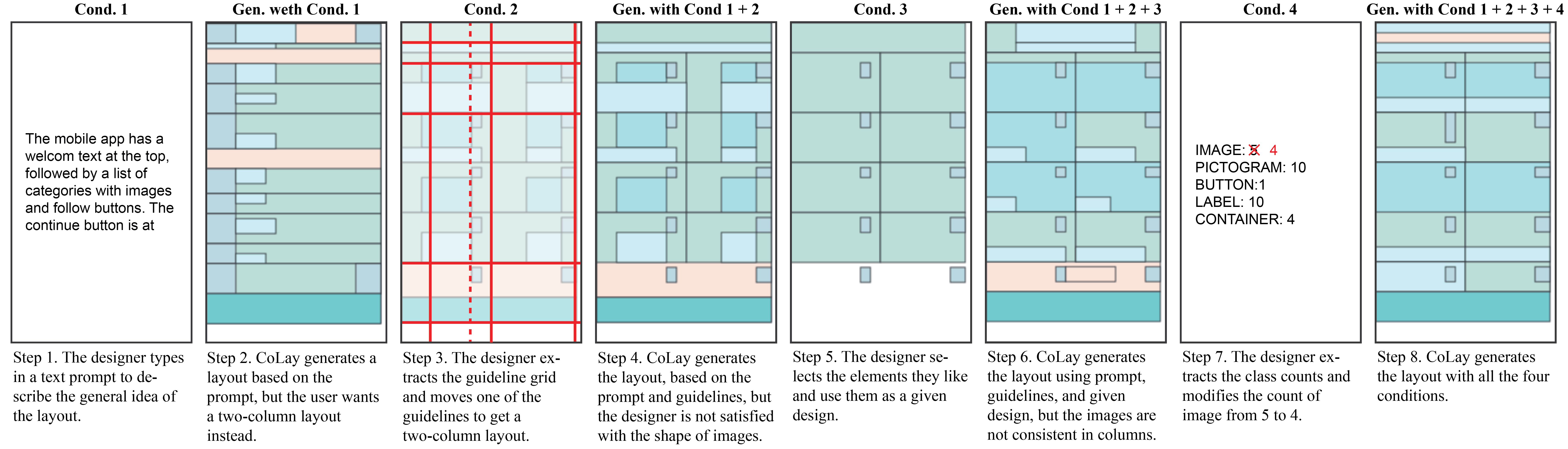}
   \caption{Multi-conditional generation and editing. We use a step-by-step example to illustrate the experience of creating and editing layouts with multiple conditions using CoLay.}
   \label{fig:seq}
\end{figure*}

In this section we demonstrate the experience of flexibly creating and editing layouts using the conditions provided by CoLay. As shown in the step-by-step examples in Figure~\ref{fig:seq}, a designer may start by describing some concepts as text prompt, leaving other conditions as empty. Then they can iterate the design by adding new conditions or modifying the conditions extracted from the generated layout, such as editing the composition of guidelines, changing the count of element classes, or regenerating a specified region.

\subsection{User Study}
Metrics such as FID scores and CLIP scores do not always reflect human preferences~\cite{kazemi2020preference}. Therefore, we also conduct user studies to verify the generation quality by human judgement. Following the user study done by~\cite{play}, we invite six designers and ask each of them to select the preferred design from 48 pairs of layouts. As shown in Table~\ref{table:user}, CoLay is preferred by designers over PLay and is comparable to ground truth samples.

\begin{table}[t]
\caption{User study results. Please refer to Appendix for details.}
\label{table:user}
\vskip 0.15in
\begin{center}
\begin{small}
% \small
% \begin{sc}
\begin{tabular}{c|ccc}
\toprule
 & G.T. & CoLay & PLay\\
\midrule
G.T. & N/A & 0.092 & 0.312 \\
CoLay & -0.092 & N/A & 0.255 \\
PLay & -0.312 & -0.255 & N/A\\
\bottomrule
\end{tabular}
% \end{sc}
\end{small}
\end{center}
\vskip -0.1in
\end{table}

\subsection{Limitations}
The layout generation quality is highly correlated to the data quality. However, both CLAY and C4 are collected years ago without applying any filters on the layout quality, and therefore, many of the layouts in these datasets are either bad designs or are outdated in terms of design styles. Also, as the text prompt pairs are screen summaries generated by LLMs, they may not be able to represent visual properties or concepts in great detail. 
% Lastly, CoLay cannot handle conflicted conditions since these situations are not included in the training set. For example, if the text prompt describes a two column layout but the given design is a single column list, one of them must be violated.
\section{Conclusion}
\label{sec:conc}
We propose CoLay, a multi-conditional latent diffusion model for layout generation trained with four commonly used conditions by designers: text prompt, class count, given design, and guidelines. We introduce new methods on both preparing and evaluating these conditions. Our model is also capable of generating style attributes and scale to complex layouts in multiple domains. CoLay not only outperforms existing methods but also provide a flexible workflow for designers to control the layout generation process by synthesising combinations of conditions.

{
    \small
    \bibliographystyle{ieeenat_fullname}
    \bibliography{main}

\begin{thebibliography}{43}
\providecommand{\natexlab}[1]{#1}
\providecommand{\url}[1]{\texttt{#1}}
\expandafter\ifx\csname urlstyle\endcsname\relax
  \providecommand{\doi}[1]{doi: #1}\else
  \providecommand{\doi}{doi: \begingroup \urlstyle{rm}\Url}\fi

\bibitem[Arroyo et~al.(2021)Arroyo, Postels, and Tombari]{vtn}
Diego~Martin Arroyo, Janis Postels, and Federico Tombari.
\newblock Variational transformer networks for layout generation.
\newblock In \emph{Proceedings of the IEEE/CVF Conference on Computer Vision
  and Pattern Recognition}, pages 13642--13652, 2021.

\bibitem[Cao et~al.(2022)Cao, Ma, Zhou, Liu, Xie, Ge, and Jiang]{icvt}
Yunning Cao, Ye Ma, Min Zhou, Chuanbin Liu, Hongtao Xie, Tiezheng Ge, and
  Yuning Jiang.
\newblock Geometry aligned variational transformer for image-conditioned layout
  generation.
\newblock In \emph{Proceedings of the 30th ACM International Conference on
  Multimedia}, pages 1561--1571, 2022.

\bibitem[Cer et~al.(2018)Cer, Yang, Kong, Hua, Limtiaco, John, Constant,
  Guajardo-Cespedes, Yuan, Tar, et~al.]{use}
Daniel Cer, Yinfei Yang, Sheng-yi Kong, Nan Hua, Nicole Limtiaco, Rhomni~St
  John, Noah Constant, Mario Guajardo-Cespedes, Steve Yuan, Chris Tar, et~al.
\newblock Universal sentence encoder.
\newblock \emph{arXiv preprint arXiv:1803.11175}, 2018.

\bibitem[Cheng et~al.(2023)Cheng, Huang, Li, and Li]{play}
Chin-Yi Cheng, Forrest Huang, Gang Li, and Yang Li.
\newblock Play: Parametrically conditioned layout generation using latent
  diffusion.
\newblock In \emph{Proceedings of the 40th International Conference on Machine
  Learning}. JMLR.org, 2023.

\bibitem[Devlin et~al.(2018)Devlin, Chang, Lee, and Toutanova]{devlin2018bert}
Jacob Devlin, Ming-Wei Chang, Kenton Lee, and Kristina Toutanova.
\newblock Bert: Pre-training of deep bidirectional transformers for language
  understanding.
\newblock \emph{arXiv preprint arXiv:1810.04805}, 2018.

\bibitem[Du et~al.(2023)Du, Durkan, Strudel, Tenenbaum, Dieleman, Fergus,
  Sohl-Dickstein, Doucet, and Grathwohl]{du2023reduce}
Yilun Du, Conor Durkan, Robin Strudel, Joshua~B Tenenbaum, Sander Dieleman, Rob
  Fergus, Jascha Sohl-Dickstein, Arnaud Doucet, and Will~Sussman Grathwohl.
\newblock Reduce, reuse, recycle: Compositional generation with energy-based
  diffusion models and mcmc.
\newblock In \emph{International Conference on Machine Learning}, pages
  8489--8510. PMLR, 2023.

\bibitem[Geffner et~al.(2022)Geffner, Papamakarios, and Mnih]{geffner2022score}
Tomas Geffner, George Papamakarios, and Andriy Mnih.
\newblock Score modeling for simulation-based inference.
\newblock In \emph{NeurIPS 2022 Workshop on Score-Based Methods}, 2022.

\bibitem[Google et~al.(2023)Google, Dai, Firat, Johnson, Lepikhin, Passos,
  Shakeri, Taropa, Bailey, Chen, Chu, Clark, Shafey, Huang, Meier-Hellstern,
  Mishra, Moreira, Omernick, Robinson, Ruder, Tay, Xiao, Xu, Zhang, Abrego,
  Ahn, Austin, Barham, Botha, Bradbury, Brahma, Brooks, Catasta, Cheng, Cherry,
  Choquette-Choo, Chowdhery, Crepy, Dave, Dehghani, Dev, Devlin, Díaz, Du,
  Dyer, Feinberg, Feng, Fienber, Freitag, Garcia, Gehrmann, Gonzalez, Gur-Ari,
  Hand, Hashemi, Hou, Howland, Hu, Hui, Hurwitz, Isard, Ittycheriah, Jagielski,
  Jia, Kenealy, Krikun, Kudugunta, Lan, Lee, Lee, Li, Li, Li, Li, Li, Lim, Lin,
  Liu, Liu, Maggioni, Mahendru, Maynez, Misra, Moussalem, Nado, Nham, Ni,
  Nystrom, Parrish, Pellat, Polacek, Polozov, Pope, Qiao, Reif, Richter, Riley,
  Ros, Roy, Saeta, Samuel, Shelby, Slone, Smilkov, So, Sohn, Tokumine, Valter,
  Vasudevan, Vodrahalli, Wang, Wang, Wang, Wang, Wieting, Wu, Xu, Xu, Xue, Yin,
  Yu, Zhang, Zheng, Zheng, Zhou, Zhou, Petrov, and Wu]{palm2}
Rohan~Anil Google, Andrew~M. Dai, Orhan Firat, Melvin Johnson, Dmitry Lepikhin,
  Alexandre Passos, Siamak Shakeri, Emanuel Taropa, Paige Bailey, Zhifeng Chen,
  Eric Chu, Jonathan~H. Clark, Laurent~El Shafey, Yanping Huang, Kathy
  Meier-Hellstern, Gaurav Mishra, Erica Moreira, Mark Omernick, Kevin Robinson,
  Sebastian Ruder, Yi Tay, Kefan Xiao, Yuanzhong Xu, Yujing Zhang,
  Gustavo~Hernandez Abrego, Junwhan Ahn, Jacob Austin, Paul Barham, Jan Botha,
  James Bradbury, Siddhartha Brahma, Kevin Brooks, Michele Catasta, Yong Cheng,
  Colin Cherry, Christopher~A. Choquette-Choo, Aakanksha Chowdhery, Clément
  Crepy, Shachi Dave, Mostafa Dehghani, Sunipa Dev, Jacob Devlin, Mark Díaz,
  Nan Du, Ethan Dyer, Vlad Feinberg, Fangxiaoyu Feng, Vlad Fienber, Markus
  Freitag, Xavier Garcia, Sebastian Gehrmann, Lucas Gonzalez, Guy Gur-Ari,
  Steven Hand, Hadi Hashemi, Le Hou, Joshua Howland, Andrea Hu, Jeffrey Hui,
  Jeremy Hurwitz, Michael Isard, Abe Ittycheriah, Matthew Jagielski, Wenhao
  Jia, Kathleen Kenealy, Maxim Krikun, Sneha Kudugunta, Chang Lan, Katherine
  Lee, Benjamin Lee, Eric Li, Music Li, Wei Li, YaGuang Li, Jian Li, Hyeontaek
  Lim, Hanzhao Lin, Zhongtao Liu, Frederick Liu, Marcello Maggioni, Aroma
  Mahendru, Joshua Maynez, Vedant Misra, Maysam Moussalem, Zachary Nado, John
  Nham, Eric Ni, Andrew Nystrom, Alicia Parrish, Marie Pellat, Martin Polacek,
  Alex Polozov, Reiner Pope, Siyuan Qiao, Emily Reif, Bryan Richter, Parker
  Riley, Alex~Castro Ros, Aurko Roy, Brennan Saeta, Rajkumar Samuel, Renee
  Shelby, Ambrose Slone, Daniel Smilkov, David~R. So, Daniel Sohn, Simon
  Tokumine, Dasha Valter, Vijay Vasudevan, Kiran Vodrahalli, Xuezhi Wang,
  Pidong Wang, Zirui Wang, Tao Wang, John Wieting, Yuhuai Wu, Kelvin Xu, Yunhan
  Xu, Linting Xue, Pengcheng Yin, Jiahui Yu, Qiao Zhang, Steven Zheng, Ce
  Zheng, Weikang Zhou, Denny Zhou, Slav Petrov, and Yonghui Wu.
\newblock Palm 2 technical report, 2023.

\bibitem[Gupta et~al.(2021)Gupta, Lazarow, Achille, Davis, Mahadevan, and
  Shrivastava]{layouttransformer}
Kamal Gupta, Justin Lazarow, Alessandro Achille, Larry~S Davis, Vijay
  Mahadevan, and Abhinav Shrivastava.
\newblock Layouttransformer: Layout generation and completion with
  self-attention.
\newblock In \emph{Proceedings of the IEEE/CVF International Conference on
  Computer Vision}, pages 1004--1014, 2021.

\bibitem[Heusel et~al.(2017)Heusel, Ramsauer, Unterthiner, Nessler, and
  Hochreiter]{fid}
Martin Heusel, Hubert Ramsauer, Thomas Unterthiner, Bernhard Nessler, and Sepp
  Hochreiter.
\newblock Gans trained by a two time-scale update rule converge to a local nash
  equilibrium.
\newblock \emph{Advances in neural information processing systems}, 30, 2017.

\bibitem[Ho and Salimans(2022)]{ho2022classifier}
Jonathan Ho and Tim Salimans.
\newblock Classifier-free diffusion guidance.
\newblock \emph{arXiv preprint arXiv:2207.12598}, 2022.

\bibitem[Ho et~al.(2020)Ho, Jain, and Abbeel]{ddpm}
Jonathan Ho, Ajay Jain, and Pieter Abbeel.
\newblock Denoising diffusion probabilistic models.
\newblock \emph{Advances in Neural Information Processing Systems},
  33:\penalty0 6840--6851, 2020.

\bibitem[Ho et~al.(2022{\natexlab{a}})Ho, Chan, Saharia, Whang, Gao, Gritsenko,
  Kingma, Poole, Norouzi, Fleet, et~al.]{ho2022imagen}
Jonathan Ho, William Chan, Chitwan Saharia, Jay Whang, Ruiqi Gao, Alexey
  Gritsenko, Diederik~P Kingma, Ben Poole, Mohammad Norouzi, David~J Fleet,
  et~al.
\newblock Imagen video: High definition video generation with diffusion models.
\newblock \emph{arXiv preprint arXiv:2210.02303}, 2022{\natexlab{a}}.

\bibitem[Ho et~al.(2022{\natexlab{b}})Ho, Saharia, Chan, Fleet, Norouzi, and
  Salimans]{ho2022cascaded}
Jonathan Ho, Chitwan Saharia, William Chan, David~J Fleet, Mohammad Norouzi,
  and Tim Salimans.
\newblock Cascaded diffusion models for high fidelity image generation.
\newblock \emph{The Journal of Machine Learning Research}, 23\penalty0
  (1):\penalty0 2249--2281, 2022{\natexlab{b}}.

\bibitem[Inoue et~al.(2023{\natexlab{a}})Inoue, Kikuchi, Simo-Serra, Otani, and
  Yamaguchi]{flexdm}
Naoto Inoue, Kotaro Kikuchi, Edgar Simo-Serra, Mayu Otani, and Kota Yamaguchi.
\newblock Towards flexible multi-modal document models.
\newblock In \emph{Proceedings of the IEEE/CVF Conference on Computer Vision
  and Pattern Recognition}, pages 14287--14296, 2023{\natexlab{a}}.

\bibitem[Inoue et~al.(2023{\natexlab{b}})Inoue, Kikuchi, Simo-Serra, Otani, and
  Yamaguchi]{inoue2023layoutdm}
Naoto Inoue, Kotaro Kikuchi, Edgar Simo-Serra, Mayu Otani, and Kota Yamaguchi.
\newblock Layoutdm: Discrete diffusion model for controllable layout
  generation.
\newblock In \emph{Proceedings of the IEEE/CVF Conference on Computer Vision
  and Pattern Recognition}, pages 10167--10176, 2023{\natexlab{b}}.

\bibitem[Jyothi et~al.(2019)Jyothi, Durand, He, Sigal, and Mori]{layoutvae}
Akash~Abdu Jyothi, Thibaut Durand, Jiawei He, Leonid Sigal, and Greg Mori.
\newblock Layoutvae: Stochastic scene layout generation from a label set.
\newblock In \emph{Proceedings of the IEEE/CVF International Conference on
  Computer Vision}, pages 9895--9904, 2019.

\bibitem[Kazemi et~al.(2020)Kazemi, Taherkhani, and
  Nasrabadi]{kazemi2020preference}
Hadi Kazemi, Fariborz Taherkhani, and Nasser Nasrabadi.
\newblock Preference-based image generation.
\newblock In \emph{Proceedings of the IEEE/CVF Winter Conference on
  Applications of Computer Vision}, pages 3404--3413, 2020.

\bibitem[Kikuchi et~al.(2021)Kikuchi, Simo-Serra, Otani, and
  Yamaguchi]{layoutganpp}
Kotaro Kikuchi, Edgar Simo-Serra, Mayu Otani, and Kota Yamaguchi.
\newblock Constrained graphic layout generation via latent optimization.
\newblock In \emph{Proceedings of the 29th ACM International Conference on
  Multimedia}, pages 88--96, 2021.

\bibitem[Kong et~al.(2022)Kong, Jiang, Chang, Zhang, Hao, Gong, and Essa]{blt}
Xiang Kong, Lu Jiang, Huiwen Chang, Han Zhang, Yuan Hao, Haifeng Gong, and
  Irfan Essa.
\newblock Blt: bidirectional layout transformer for controllable layout
  generation.
\newblock In \emph{Computer Vision--ECCV 2022: 17th European Conference, Tel
  Aviv, Israel, October 23--27, 2022, Proceedings, Part XVII}, pages 474--490.
  Springer, 2022.

\bibitem[Lee et~al.(2020)Lee, Jiang, Essa, Le, Gong, Yang, and
  Yang]{neuraldesign}
Hsin-Ying Lee, Lu Jiang, Irfan Essa, Phuong~B Le, Haifeng Gong, Ming-Hsuan
  Yang, and Weilong Yang.
\newblock Neural design network: Graphic layout generation with constraints.
\newblock In \emph{European Conference on Computer Vision}, pages 491--506.
  Springer, 2020.

\bibitem[Li et~al.(2022)Li, Baechler, Tragut, and Li]{clay}
Gang Li, Gilles Baechler, Manuel Tragut, and Yang Li.
\newblock Learning to denoise raw mobile ui layouts for improving datasets at
  scale.
\newblock In \emph{CHI Conference on Human Factors in Computing Systems}, pages
  1--13, 2022.

\bibitem[Li et~al.(2019)Li, Yang, Hertzmann, Zhang, and Xu]{layoutgan}
Jianan Li, Jimei Yang, Aaron Hertzmann, Jianming Zhang, and Tingfa Xu.
\newblock Layoutgan: Generating graphic layouts with wireframe discriminators.
\newblock \emph{arXiv preprint arXiv:1901.06767}, 2019.

\bibitem[Li et~al.(2020)Li, Yang, Zhang, Liu, Wang, and Xu]{attribute}
Jianan Li, Jimei Yang, Jianming Zhang, Chang Liu, Christina Wang, and Tingfa
  Xu.
\newblock Attribute-conditioned layout gan for automatic graphic design.
\newblock \emph{IEEE Transactions on Visualization and Computer Graphics},
  27\penalty0 (10):\penalty0 4039--4048, 2020.

\bibitem[Lin et~al.(2023)Lin, Guo, Sun, Yang, Lou, and
  Zhang]{lin2023layoutprompter}
Jiawei Lin, Jiaqi Guo, Shizhao Sun, Zijiang~James Yang, Jian-Guang Lou, and
  Dongmei Zhang.
\newblock Layoutprompter: Awaken the design ability of large language models.
\newblock \emph{arXiv preprint arXiv:2311.06495}, 2023.

\bibitem[Liu et~al.(2018)Liu, Craft, Situ, Yumer, Mech, and
  Kumar]{ricosemantic}
Thomas~F Liu, Mark Craft, Jason Situ, Ersin Yumer, Radomir Mech, and Ranjitha
  Kumar.
\newblock Learning design semantics for mobile apps.
\newblock In \emph{Proceedings of the 31st Annual ACM Symposium on User
  Interface Software and Technology}, pages 569--579, 2018.

\bibitem[Nair et~al.(2022)Nair, Bandara, and Patel]{nair2022unite}
Nithin~Gopalakrishnan Nair, Wele Gedara~Chaminda Bandara, and Vishal~M Patel.
\newblock Unite and conquer: Cross dataset multimodal synthesis using diffusion
  models.
\newblock \emph{arXiv preprint arXiv:2212.00793}, 2022.

\bibitem[Nauata et~al.(2021)Nauata, Hosseini, Chang, Chu, Cheng, and
  Furukawa]{houseganp}
Nelson Nauata, Sepidehsadat Hosseini, Kai-Hung Chang, Hang Chu, Chin-Yi Cheng,
  and Yasutaka Furukawa.
\newblock House-gan++: Generative adversarial layout refinement network towards
  intelligent computational agent for professional architects.
\newblock In \emph{Proceedings of the IEEE/CVF Conference on Computer Vision
  and Pattern Recognition}, pages 13632--13641, 2021.

\bibitem[Nguyen et~al.(2021)Nguyen, Nepal, and Kanhere]{mcl}
David~D Nguyen, Surya Nepal, and Salil~S Kanhere.
\newblock Diverse multimedia layout generation with multi choice learning.
\newblock In \emph{Proceedings of the 29th ACM International Conference on
  Multimedia}, pages 218--226, 2021.

\bibitem[Perez et~al.(2018)Perez, Strub, De~Vries, Dumoulin, and
  Courville]{film}
Ethan Perez, Florian Strub, Harm De~Vries, Vincent Dumoulin, and Aaron
  Courville.
\newblock Film: Visual reasoning with a general conditioning layer.
\newblock In \emph{Proceedings of the AAAI Conference on Artificial
  Intelligence}, 2018.

\bibitem[Raffel et~al.(2020)Raffel, Shazeer, Roberts, Lee, Narang, Matena,
  Zhou, Li, and Liu]{2020t5}
Colin Raffel, Noam Shazeer, Adam Roberts, Katherine Lee, Sharan Narang, Michael
  Matena, Yanqi Zhou, Wei Li, and Peter~J. Liu.
\newblock Exploring the limits of transfer learning with a unified text-to-text
  transformer.
\newblock \emph{Journal of Machine Learning Research}, 21\penalty0
  (140):\penalty0 1--67, 2020.

\bibitem[Ramesh et~al.(2022)Ramesh, Dhariwal, Nichol, Chu, and
  Chen]{ramesh2022hierarchical}
Aditya Ramesh, Prafulla Dhariwal, Alex Nichol, Casey Chu, and Mark Chen.
\newblock Hierarchical text-conditional image generation with clip latents.
\newblock \emph{arXiv preprint arXiv:2204.06125}, 1\penalty0 (2):\penalty0 3,
  2022.

\bibitem[Saharia et~al.(2022{\natexlab{a}})Saharia, Chan, Chang, Lee, Ho,
  Salimans, Fleet, and Norouzi]{saharia2022palette}
Chitwan Saharia, William Chan, Huiwen Chang, Chris Lee, Jonathan Ho, Tim
  Salimans, David Fleet, and Mohammad Norouzi.
\newblock Palette: Image-to-image diffusion models.
\newblock In \emph{ACM SIGGRAPH 2022 Conference Proceedings}, pages 1--10,
  2022{\natexlab{a}}.

\bibitem[Saharia et~al.(2022{\natexlab{b}})Saharia, Chan, Saxena, Li, Whang,
  Denton, Ghasemipour, Ayan, Mahdavi, Lopes, et~al.]{imagen}
Chitwan Saharia, William Chan, Saurabh Saxena, Lala Li, Jay Whang, Emily
  Denton, Seyed Kamyar~Seyed Ghasemipour, Burcu~Karagol Ayan, S~Sara Mahdavi,
  Rapha~Gontijo Lopes, et~al.
\newblock Photorealistic text-to-image diffusion models with deep language
  understanding.
\newblock \emph{arXiv preprint arXiv:2205.11487}, 2022{\natexlab{b}}.

\bibitem[Saharia et~al.(2022{\natexlab{c}})Saharia, Chan, Saxena, Li, Whang,
  Denton, Ghasemipour, Gontijo~Lopes, Karagol~Ayan, Salimans,
  et~al.]{saharia2022photorealistic}
Chitwan Saharia, William Chan, Saurabh Saxena, Lala Li, Jay Whang, Emily~L
  Denton, Kamyar Ghasemipour, Raphael Gontijo~Lopes, Burcu Karagol~Ayan, Tim
  Salimans, et~al.
\newblock Photorealistic text-to-image diffusion models with deep language
  understanding.
\newblock \emph{Advances in Neural Information Processing Systems},
  35:\penalty0 36479--36494, 2022{\natexlab{c}}.

\bibitem[Szegedy et~al.(2017)Szegedy, Ioffe, Vanhoucke, and Alemi]{inception}
Christian Szegedy, Sergey Ioffe, Vincent Vanhoucke, and Alexander~A Alemi.
\newblock Inception-v4, inception-resnet and the impact of residual connections
  on learning.
\newblock In \emph{Thirty-first AAAI conference on artificial intelligence},
  2017.

\bibitem[Wang et~al.(2023)Wang, Li, and Li]{bryan}
Bryan Wang, Gang Li, and Yang Li.
\newblock Enabling conversational interaction with mobile {UI} using large
  language models.
\newblock In \emph{Proceedings of the 2023 {CHI} Conference on Human Factors in
  Computing Systems, {CHI} 2023, Hamburg, Germany, April 23-28, 2023}, pages
  432:1--432:17. {ACM}, 2023.

\bibitem[Xu et~al.(2022)Xu, Willis, Lambourne, Cheng, Jayaraman, and
  Furukawa]{skexgen}
Xiang Xu, Karl~DD Willis, Joseph~G Lambourne, Chin-Yi Cheng, Pradeep~Kumar
  Jayaraman, and Yasutaka Furukawa.
\newblock Skexgen: Autoregressive generation of cad construction sequences with
  disentangled codebooks.
\newblock \emph{arXiv preprint arXiv:2207.04632}, 2022.

\bibitem[Yamaguchi(2021)]{canvasvae}
Kota Yamaguchi.
\newblock Canvasvae: Learning to generate vector graphic documents.
\newblock In \emph{Proceedings of the IEEE/CVF International Conference on
  Computer Vision}, pages 5481--5489, 2021.

\bibitem[Zhang et~al.(2023)Zhang, Rao, and Agrawala]{controlnet}
Lvmin Zhang, Anyi Rao, and Maneesh Agrawala.
\newblock Adding conditional control to text-to-image diffusion models.
\newblock In \emph{Proceedings of the IEEE/CVF International Conference on
  Computer Vision}, pages 3836--3847, 2023.

\bibitem[Zhong et~al.(2019)Zhong, Tang, and Yepes]{publaynet}
Xu Zhong, Jianbin Tang, and Antonio~Jimeno Yepes.
\newblock Publaynet: largest dataset ever for document layout analysis.
\newblock In \emph{2019 International Conference on Document Analysis and
  Recognition (ICDAR)}, pages 1015--1022. IEEE, 2019.

\bibitem[Zhou et~al.(2022)Zhou, Xu, Ma, Ge, Jiang, and Xu]{cglgan}
Min Zhou, Chenchen Xu, Ye Ma, Tiezheng Ge, Yuning Jiang, and Weiwei Xu.
\newblock Composition-aware graphic layout gan for visual-textual presentation
  designs.
\newblock \emph{arXiv preprint arXiv:2205.00303}, 2022.

\bibitem[Zhu et~al.(2017)Zhu, Park, Isola, and Efros]{cycle}
Jun-Yan Zhu, Taesung Park, Phillip Isola, and Alexei~A Efros.
\newblock Unpaired image-to-image translation using cycle-consistent
  adversarial networks.
\newblock In \emph{Proceedings of the IEEE international conference on computer
  vision}, pages 2223--2232, 2017.

\end{thebibliography}
}

\clearpage
\setcounter{page}{1}
\maketitlesupplementary

\section{Additional Statistics on Datasets}
\label{sec:data_details}

\begin{figure*}[t]
  \centering
   \includegraphics[width=1.0\linewidth]{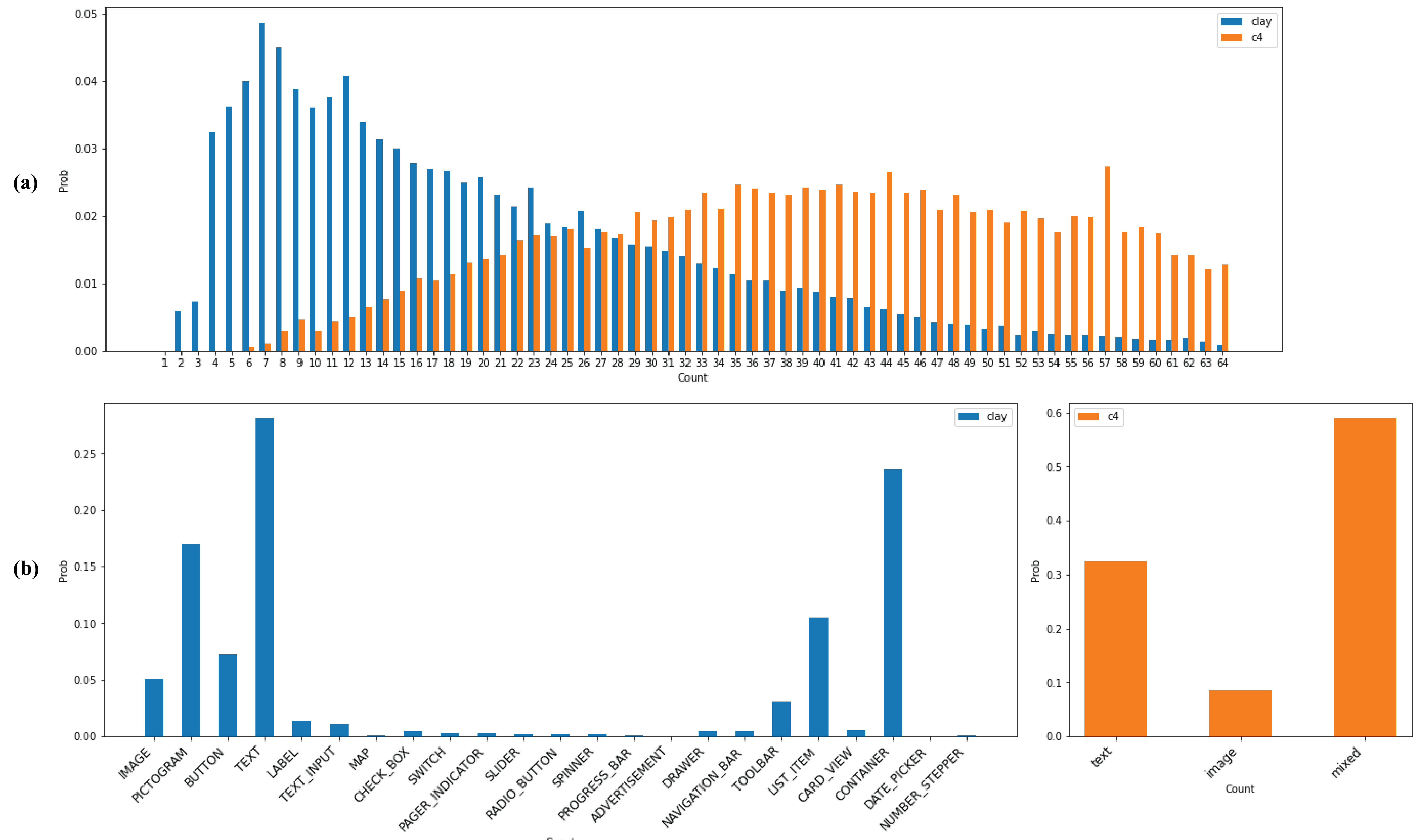}

   \caption{Dataset statistics. (a) Number of element distribution for CLAY and C4, where the x-axis represents the number of elements in a layout and the y-axis is the corresponding probability. (b) Left: type distribution for CLAY; right: type distribution for C4.}
   \label{fig:dataset}
\end{figure*}

We provide additional statistics on the CLAY and C4 datasets in Figure~\ref{fig:dataset}.

\section{Sampling with Condition Combinations}
\label{sec:comb}

In Table~\ref{table:comb_clay} and \ref{table:comb_c4}, we demonstrate how CoLay trained on all four conditions performs when generating samples using different number of conditions and different combinations of conditions. Interestingly, we observe that in some condition combinations, the four-condition model can outperform the specialized models by searching the optimal classifier-free guidance weights. We also notice that in Table~\ref{table:comb_c4}, the FID scores decrease much more on the C4 dataset when generating layouts with less number of conditions. For example, the FID score drops from 10.66 to XXXX when using the class condition alone.

\begin{table}[t]
\caption{FID scores in all different condition combinations on the CLAY dataset. In each row, the weights represent the optimal weights for this combination of conditions. Single and Multi in every row use the same model that is trained on four conditions. Single: using a single guidance weight, $2.5$, for all conditions. Multi: using different weights for each condition and search for optimal weights. Specialized: a specialized model for this condition combination (row). Note that we only include specialized models for some combinations due to limited time and resource.}
\label{table:comb_clay}
\vskip 0.15in
\begin{center}
\begin{small}
% \small
% \begin{sc}
\begin{tabular}{cccc|ccc}
\toprule
\multicolumn{7}{c}{\textbf{Dataset: CLAY}}\\
\midrule
\multicolumn{4}{c|}{Combinations/Weights} & \multicolumn{3}{|c}{FID} \\
\cmidrule(r){1-4}\cmidrule(r){5-7}
$G$ & $E^\prime$ & $C$ & $P$ & Single & Multi & Specialized\\
\midrule
1.8 & 1.3 & 1.9 & 2.3 & 6.54 & \textbf{6.23} & \textbf{6.23}\\
1.9 & 2.0 & & 2.4 & 6.73 & 6.49 & ---\\
& 2.2 & 1.9 & 2.4 & 6.81 & 6.72 & ---\\
2.4 & 2.2 & 2.2 & & 6.85 & \textbf{6.62} & 6.94\\
& 2.1 & 2.7 & & 6.92 & 6.67 & ---\\
2.6 & 2.3 & & & 7.37 & 7.28 & \textbf{7.23}\\
& 2.5 & & 3.3 & 7.58 & 7.18 & ---\\
2.2 & & 3.7 & 2.3 & 8.48 & \textbf{8.19} & 8.22\\
2.3 & & 3.7 & & 8.67 & 8.41 & ---\\
& & 3.6 & 3.6 & 8.79 & \textbf{8.49} & 8.54\\
2.1 & & & 3.8 & 9.17 & 8.66 & \textbf{8.42}\\
& 2.6 & & & 9.41 & 9.13 & ---\\
3.3 & & & & 9.76 & 9.69 & \textbf{8.77}\\
& & 3.8 & & 10.65 & 10.09 & ---\\
& & & 3.4 & 10.95 & 9.82 & \textbf{8.93}\\
\bottomrule
\end{tabular}
% \end{sc}
\end{small}
\end{center}
\vskip -0.1in
\end{table}

\begin{table}[t]
\caption{FID scores in all different condition combinations on the C4 dataset. In each row, the weights represent the optimal weights for this combination of conditions. Single and Multi in every row use the same model that is trained on four conditions. Single: using a single guidance weight, $2.5$, for all conditions. Multi: using different weights for each condition and search for optimal weights. Specialized: a specialized model for this condition combination (row). Note that we only include specialized models for some combinations due to limited time and resource}
\label{table:comb_c4}
\vskip 0.15in
\begin{center}
\begin{small}
% \small
% \begin{sc}
\begin{tabular}{cccc|ccc}
\toprule
\multicolumn{7}{c}{\textbf{Dataset: C4}}\\
\midrule
\multicolumn{4}{c|}{Combinations/Weights} & \multicolumn{3}{|c}{FID} \\
\cmidrule(r){1-4}\cmidrule(r){5-7}
$G$ & $E^\prime$ & $C$ & $P$ & Single & Multi & Specialized\\
\midrule

2.1 & 2.0 & 1.9 & 2.4 & 11.21 & \textbf{10.66} & \textbf{10.66} \\
2.2 & 2.2 & 2.4 & & 11.02 & \textbf{10.75} & 12.45 \\
2.1 & 2.0 & & 2.1 & 11.15 & 10.73 & --- \\
2.2 & 1.8 & & & 11.17 & 10.69 & ---\\
2.2 & & 3.3 & & 14.18 & 14.09 & ---\\
2.2 & & & 3.7 & 14.21 & 13.85 & ---\\
& 2.1 & 3.4 & & 14.34 & 13.61 & ---\\
& 1.8 & 2.6 & 2.5 & 14.38 & 13.61 & ---\\
2.8 & & & & 14.52 & 14.33 & ---\\
& 2.1 & & 3.5 & 14.54 & 14.07 & ---\\
2.3 & & 3.3 & 2.6 & 14.56 & 14.12 & ---\\
& 2.9 & & & 19.50 & 19.14 & ---\\
& & & 2.6 & 24.54 & 24.36 & ---\\
& & 3.1 & 3.3 & 24.37 & 24.06 & ---\\
& & 3.6 & & 27.06 & 25.12 & ---\\
\bottomrule
\end{tabular}
% \end{sc}
\end{small}
\end{center}
\vskip -0.1in
\end{table}

\section{Implementation Details}
\label{sec:imp}
The denoiser we use for CoLay, PLay, and LayoutDM in this paper is a transformer encoder that has $4$ layers and $8$ heads with $512$ for the QKV dimension and $2048$ for the MLP dimension. The latent dimension $\hat{d}$ for $z$ is $8$ for CLAY and $32$ for C4. We also increase the QKV dimension to $1024$ when training on C4. Both the guideline encoder and class count encoder share similar architecture as the denoiser with the number of layers decreased to $2$.

CoLay is implemented using \textit{JAX} and \textit{Flax} and trained using 16 Google Cloud TPU v3 cores with batch size of $1024$. The number of training steps for CLAY and C4 are $500k$ and $1.5m$, respectively. We use ADAM optimizer with $b_1=0.9$, $b_2=0.98$.

\section{UI Summary Task using LLMs}
\label{sec:llm_summary}
Since there are no human-annotated text captions associated with the C4 and CLAY datasets, and the only publically available dataset of text-UI pairs are relatively small in its scale, we use an LLM based on the PaLM-2-Bison model to obtain text prompts for C4 and CLAY datasets.

Given a UI from C4 or CLAY, we first flatten it and extract the interactable elements from them. This usually represent elements that has content in it, or are clickable, or are of an interactable class (i.e., non-container classes in CLAY). Then, we represent this set of elements in a flattened HTML format. The content of the elements and its type are represented in the HTML. For non-text contents, we utilize the alt-text from the UI as the content representation of them. As an example, a login page with a text instruction of ``Enter your crednetials below'', two text fields for username and password, and a ``Login'' button can be represented as follows:

\begin{verbatim}
<screen>
<p>Enter your credentials below:</p>
<input>Username</input>
<input>Password</input>
<button>Login</button>
</screen>
\end{verbatim}

This representation follows prior work that utilizes LLMs for UI-based interactions~\cite{bryan}. The HTML representation of the UIs are then combined with various types of prompt text to obtain various types of summaries of the UIs. These summaries focus on the layout, functionality, usability, and a high-level overview of the UIs. The prompts are listed as follows.

Layout prompt:
\begin{verbatim}
Below is a simplified HTML code of a
(mobile app / website):

<screen> ... (screen of the UI) </screen>

Briefly describe the layout of the
(mobile app / website) in a few sentences. 
\end{verbatim}

Functionality prompt:
\begin{verbatim}
Below is a simplified HTML code of a
(mobile app / website):

<screen> ... (screen of the UI) </screen>

Describe the functionality of the
(mobile app / website).
\end{verbatim}

Usability prompt:
\begin{verbatim}
Below is a simplified HTML code of a
(mobile app / website):

<screen> ... (screen of the UI) </screen>

What can a user do with the
(mobile app / website)?
\end{verbatim}

High-level overview prompt:
\begin{verbatim}
Below is a simplified HTML code of a
(mobile app / website):

<screen> ... (screen of the UI) </screen>

Summarize the UI of the
(mobile app / website).
\end{verbatim}

The results of after passing these prompts for LLM inference are treated as text captions of the UIs in CoLay, for further training. Figure~\ref{fig:prompts} shows an example of a UI layout paired with the four types of summaries listed above:

\begin{figure*}[t]
  \centering
   \includegraphics[width=0.9\linewidth]{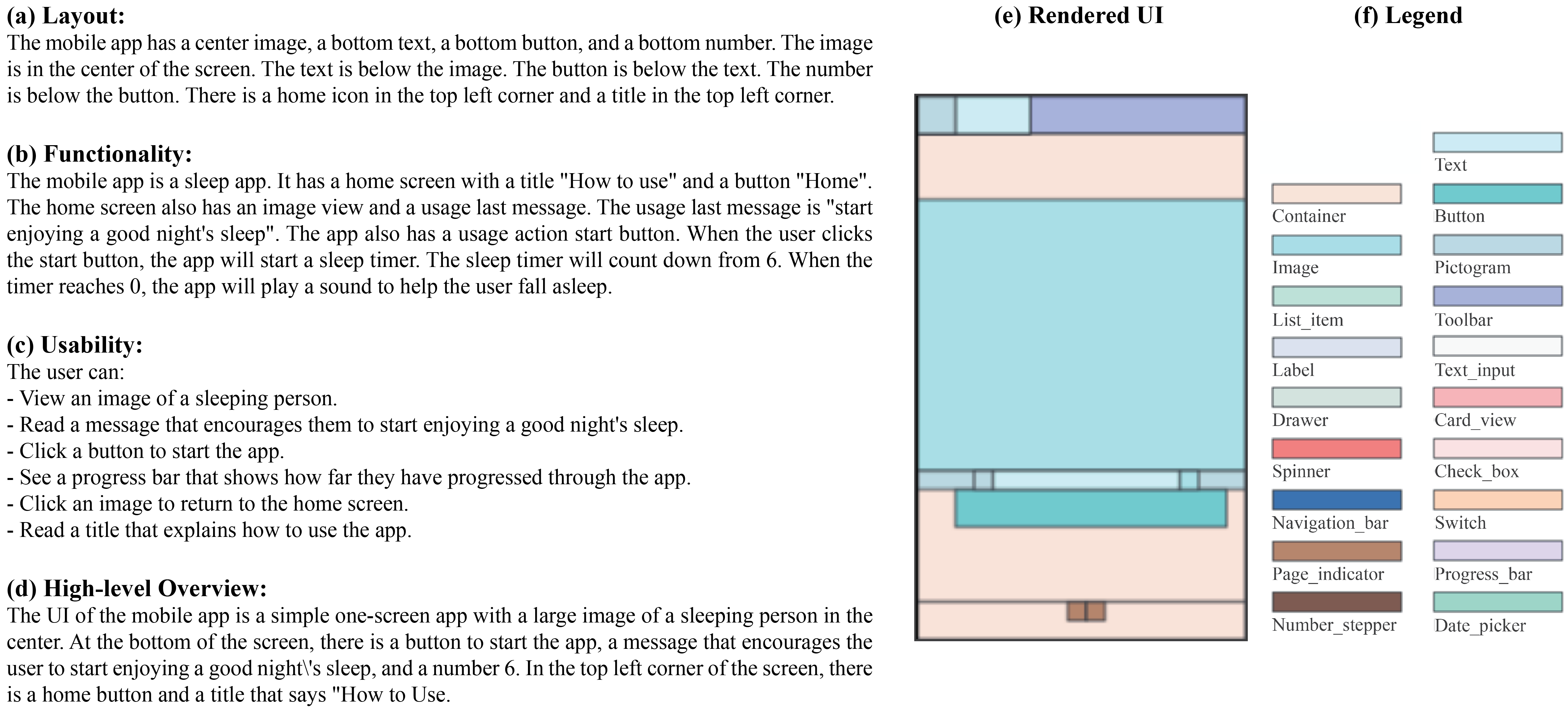}

   \caption{Example UI layout paired with four types of generated UI summaries. From (a) to (d): generated summaries with different prompts. (e) Rendered input HTML. (f) Color legend.}
   \label{fig:prompts}
\end{figure*}

From the experiment results in Table~\ref{table:prompt_fid}, we find that the model trained on the layout description prompts performs the best on generation quality (FID) and text-layout alignment (CycSim). The potential reason is that all the three other prompt types contain information about the content, which is less relevant to the layout elements. For example, the UI layout will remain the same no matter the image is a sleeping person or a bed. The rank of the performance among these prompt types may change if we include content, such as article title and images, as input conditions in the future.

\begin{table}[t]
\caption{Quantitative Results on the CLAY dataset using different types of prompts. Note that the models in this table is conditioned on text prompts only.}
\label{table:prompt_fid}
\vskip 0.15in
\begin{center}
\begin{small}
% \small
% \begin{sc}
\begin{tabular}{l|cccccc}
\toprule
Prompt Type & FID & CycSim-L & CycSim-P\\
\midrule
Layout & \textbf{8.93} & \textbf{0.779} & \textbf{0.506}\\
Functionality & 9.00 & 0.767 & 0.409\\
Usability & 9.44 & 0.769 & 0.304\\
High-level Overview & 8.95 & 0.770 & 0.459\\
\bottomrule
\end{tabular}
% \end{sc}
\end{small}
\end{center}
\vskip -0.1in
\end{table}

\section{User Study Details}
\label{sec:user_details}
We follow the user study setup in ~\cite{play, houseganp} and invite six designers who have experience on UI design. We prepare 96 questions using three groups of randomly selected layouts: layouts generated by CoLay, layouts generated by PLay, and the ground truth layouts in CLAY. For each question we pick two layouts from different groups and ask the participant to choose the better one based on the design quality, and we ensure each question is answered by three different participants. When computing the results for each question, the group will receive +1 score if it is preferred by the designer, -1 vice versa. Both groups will receive 0 if the designer thinks they are equally good or bad. The final score each group-to-group comparison is normalized by the number of questions in that category.

\section{Additional Qualitative Results}
\label{sec:add_qualitative}

\begin{figure*}[t]
  \centering
   \includegraphics[width=0.68\linewidth]{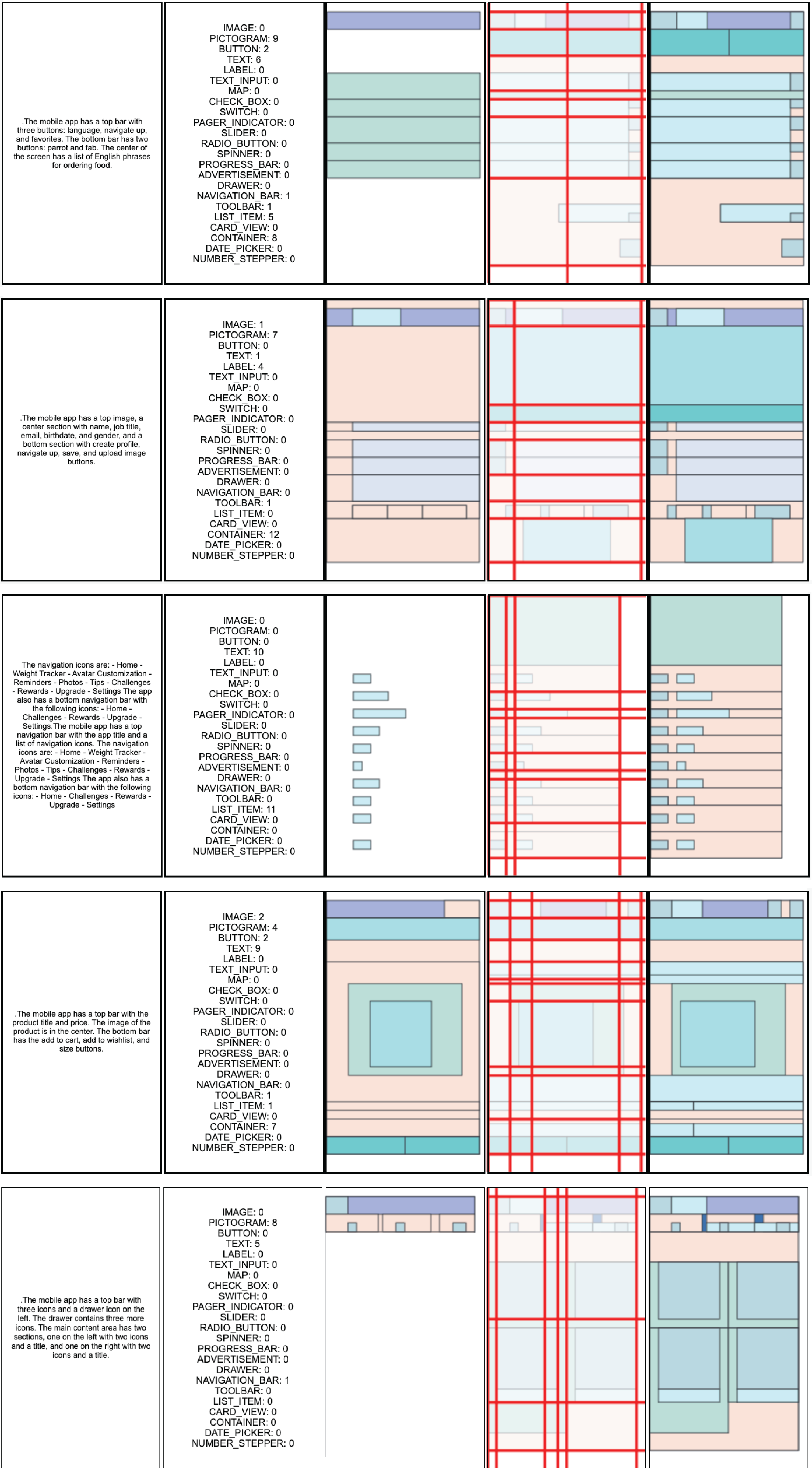}

   \caption{CLAY results with full conditions." }
   \label{fig:clay_more}
\end{figure*}

\begin{figure*}[t]
  \centering
   \includegraphics[width=0.68\linewidth]{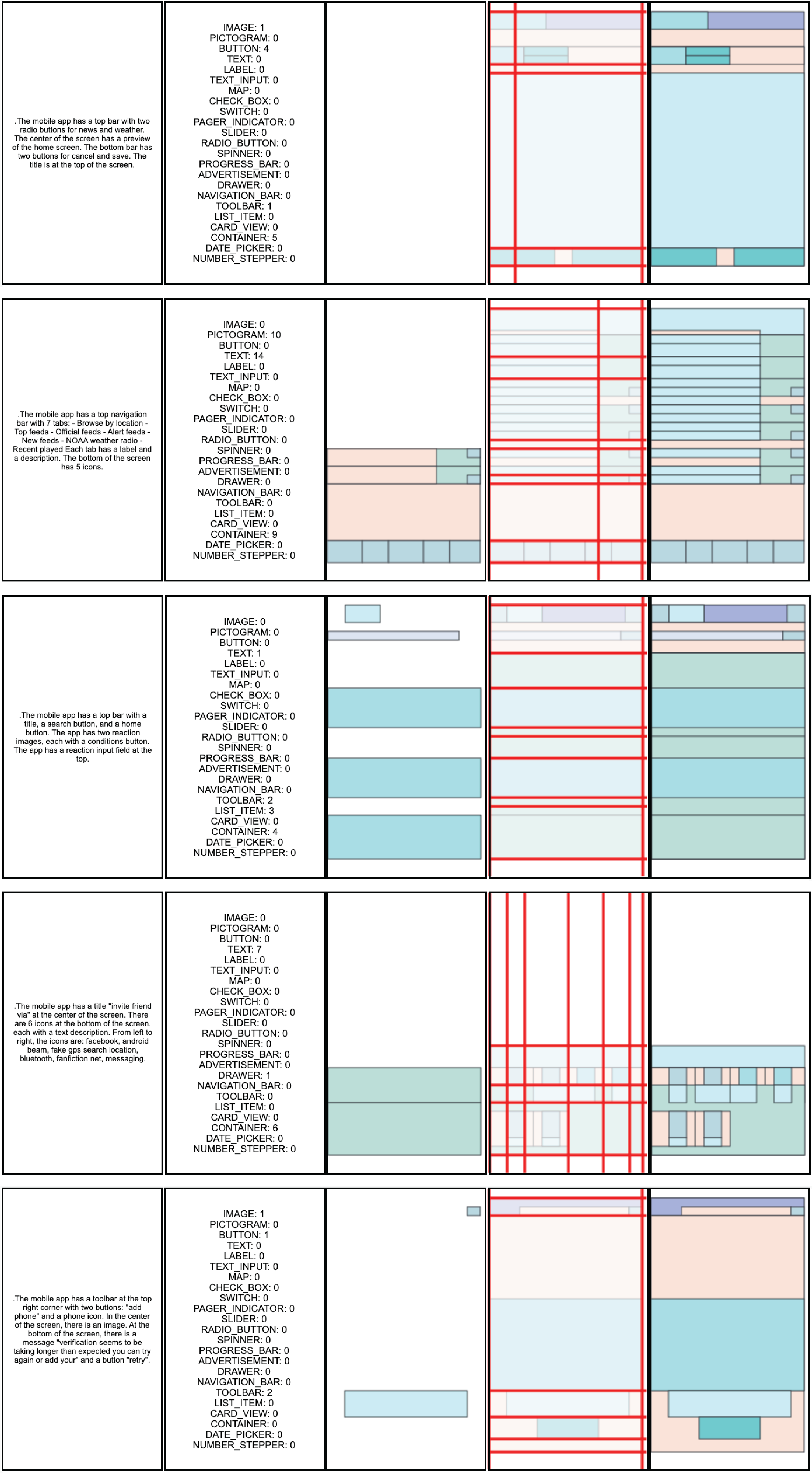}

   \caption{(continued) CLAY results with full conditions.}
   \label{fig:clay_more2}
\end{figure*}

\begin{figure*}[t]
  \centering
   \includegraphics[width=1.0\linewidth]{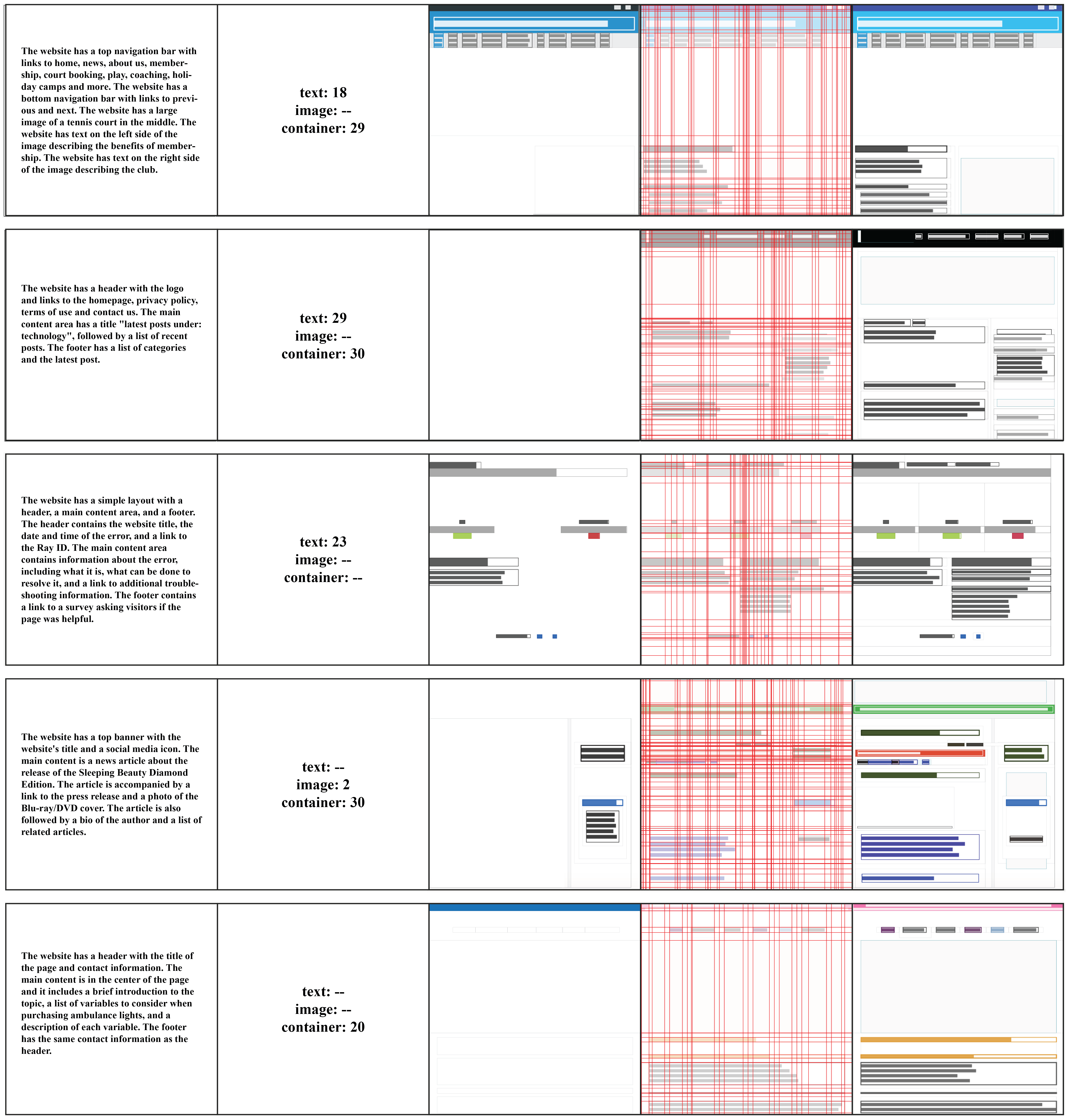}

   \caption{C4 results with full conditions. Note that we use dummy text to represent font-size and text alignment and represent font-weight by the thickness of the boarders. "--" in the type and count condition here represents "not given." }
   \label{fig:c4_more}
\end{figure*}

\begin{figure*}[t]
  \centering
   \includegraphics[width=1.0\linewidth]{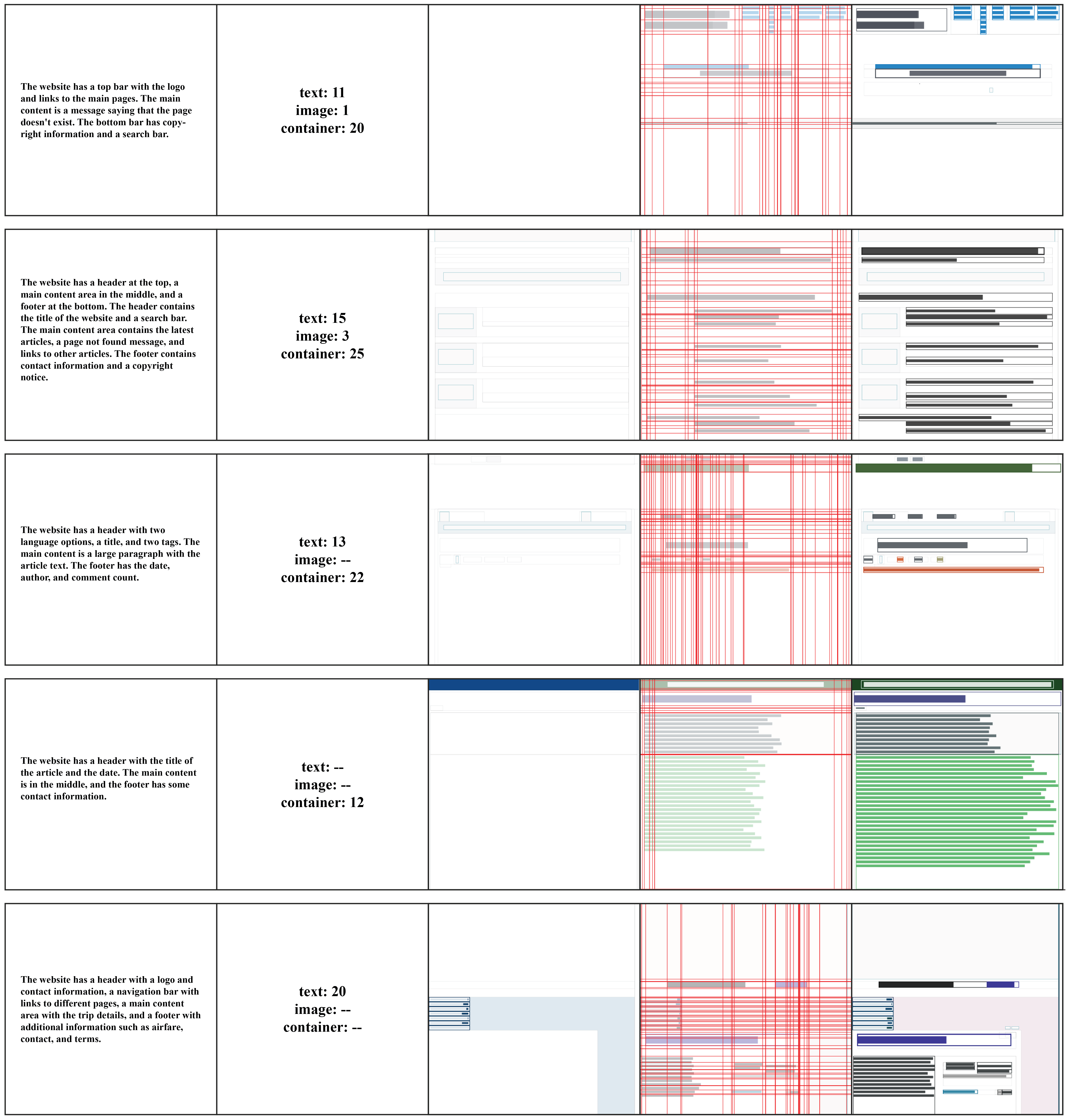}

   \caption{(continued) C4 results with full conditions.}
   \label{fig:c4_more2}
\end{figure*}

\end{document}